\documentstyle[12pt]{article}
\catcode`\@=11
\def\marginnote#1{}
\newcount\hour
\newcount\minute
\newtoks\amorpm
\hour=\time\divide\hour by60
\minute=\time{\multiply\hour by60 \global\advance\minute by-\hour}
\edef\standardtime{{\ifnum\hour<12 \global\amorpm={am}%
        \else\global\amorpm={pm}\advance\hour by-12 \fi
        \ifnum\hour=0 \hour=12 \fi
        \number\hour:\ifnum\minute<10 0\fi\number\minute\the\amorpm}}
\edef\militarytime{\number\hour:\ifnum\minute<10 0\fi\number\minute}

\def\draftlabel#1{{\@bsphack\if@filesw {\let\thepage\relax
   \xdef\@gtempa{\write\@auxout{\string
      \newlabel{#1}{{\@currentlabel}{\thepage}}}}}\@gtempa
   \if@nobreak \ifvmode\nobreak\fi\fi\fi\@esphack}
        \gdef\@eqnlabel{#1}}
\def\@eqnlabel{}
\def\@vacuum{}
\def\draftmarginnote#1{\marginpar{\raggedright\scriptsize\tt#1}}

\def\draft{\oddsidemargin -.5truein
        \def\@oddfoot{\sl preliminary draft \hfil
        \rm\thepage\hfil\sl\today\quad\militarytime}
        \let\@evenfoot\@oddfoot \overfullrule 3pt
        \let\label=\draftlabel
        \let\marginnote=\draftmarginnote
   \def\@eqnnum{(\theequation)\rlap{\kern\marginparsep\tt\@eqnlabel}%
\global\let\@eqnlabel\@vacuum}  }

\def\numberbysection{\@addtoreset{equation}{section}
        \def\theequation{\thesection.\arabic{equation}}}

\def\underline#1{\relax\ifmmode\@@underline#1\else
        $\@@underline{\hbox{#1}}$\relax\fi}

\def\titlepage{\@restonecolfalse\if@twocolumn\@restonecoltrue\onecolumn
     \else \newpage \fi \thispagestyle{empty}\c@page\z@
        \def\thefootnote{\fnsymbol{footnote}} }

\def\endtitlepage{\if@restonecol\twocolumn \else  \fi
        \def\thefootnote{\arabic{footnote}}
        \setcounter{footnote}{0}}  %\c@footnote\z@ }
\catcode`@=12
\relax

\numberbysection
\def\beq{\begin{equation}}
\def\eeq{\end{equation}}
\newtheorem{th}{Theorem}[section]
\newtheorem{cor}{Corollary}[section]

\begin{document}
\begin{titlepage}

\nopagebreak
\begin{flushright}

LPTENS--92/18,\\
hep-th@xxx/92XXXX \\
                May 1992
\end{flushright}

\vglue 2.5  true cm
\begin{center}
{\large\bf
The $\tau$-function of the universal Whitham hierarchy,\\
 matrix models and topological field theories}

\vglue 1  true cm

{\bf I.M.Krichever}
\footnote{ On leave of absence from Landau Institute
for Theoretical Physics,\\
Kosygina str. 2, 117940 Moscow, Russian Federation}\\
{\footnotesize Laboratoire de Physique Th\'eorique de
l'\'Ecole Normale Sup\'erieure\footnote{Unit\'e Propre du
Centre National de la Recherche Scientifique,
associ\'ee \`a l'\'Ecole Normale Sup\'erieure et \`a l'Universit\'e
de Paris-Sud.},\\
24 rue Lhomond, 75231 Paris CEDEX 05, ~France.}
\end{center}
\vfill
\begin{abstract}
The universal Witham hierarchy is considered from the point of view
of topological field theories. The $\tau$-function for this hierarchy is
defined. It is proved that the algebraic orbits of Whitham hierarchy
can be identified with various topological matter models coupled with
topological gravity.
\end{abstract}
\vfill

\end{titlepage}

\section{Introduction}

The last two years breakthrough in low-dimensional string theory is one of
most exiting results in modern mathematical and theoretical physics. In
\cite{BK..} remarkable connections between the non-perturbative theory
of two-dimensional gravity coupled with various matter fields \cite{npergr}
and the theory of integrable KdV-type systems were found that had led
to complete solvability of double-scaling limit of the matrix-models
that are used to simulate fluctuating triangulated Riemann surfaces.
Shortly after that Witten \cite{W1} presented some evidence for a
relationship between random surfaces and the algebraic topology of moduli
spaces of Riemann surfaces with punctures. His approach involved
a particular field theory, known as topological gravity \cite{TopG}.
Further developments of his approach ( especially, Kontsevich's
proof \cite {Kon} of Witten's conjecture \cite{W1,W2} on the
coincidence of the generating function for intersection numbers
of moduli spaces with $\tau$-function of the KdV hierarchy) have shown that
two-dimensional topological gravity is the corestone of this new subject
of mathematical physics that includes: two-dimensional quantum field theories,
intersection theory on the moduli spaces of Riemann surfaces with punctures,
integrable hierarchies with special Virasoro-type constraints, matrix
integrals,
random surfaces and so on.

In this paper we continue our previous attempts \cite{K1,K2} to find
a right place in this range of disciplines for the Whitham theory
which is a most interesting part of perturbation theory of KdV-type integrable
hierarchies. They were stimulated by results of \cite{VVD},
where correlation functions for topological minimal models were found.
It turned out that calculations \cite{VVD} of perturbed primary rings for
$A_n$-models can be identified with the construction of a particular
solution of the first $n$ ``flows" of the dispersionless Lax
hierarchy (semi-classical limit of the usual Lax hierarchy). It provided
a possibility to include corresponding deformations of primary chiral rings
into a hierarchy of inifinite number of commuting ``flows".
The calculations \cite{VVD}
of partition function for perturbed $A_n$ models gave an impulse for
introduction in \cite{K1} a $\tau$-function for dispersionless Lax
equations. The truncated version of Virasoro constraints for the
corresponding $\tau$-function were proved. Their comparision with
\cite{W1} shows that they are necessary conditions for identification of
``generators" of ``higher" flows with gravitational descendants of primary
fields after coupling model with gravity. For $n>2$ they are not sufficient.
(The problem is the same as for $\tau$-function of multi-matrix models.
As it was shown in \cite{FK} the $\tau$-function of in multi-matrix models
satisfies and uniquelly defined with the help of higher $W$-constraints.) In
section 4 we prove the truncated version of $W$-constraints for $\tau$-function
of dispersionless Lax equations. Therefore, the full dispersionless Lax
hierarchy really can be identified with topological $A_n$
minimal model coupled with gravity.

The results of \cite{K1} were generalized in \cite{K2,D2} for higher
genus case. In \cite{K2} it was shown that self-similar solutions of the
Whitham equations on the moduli space of genus $g$ Riemann surfaces are
related with ``multi-cut" solutions of loop-equations for matrix models.
In \cite{D2} the generalization of topological
Landau-Ginsburg models on Riemann surfaces of special type were proposed and
their primary rings and correlation function were found.
In \cite{D2} the Hamiltonian formulation \cite{ND} of the Whitham averaging
procedure was used. With its help it was proved that ``coupling"
constants for primary fields of such models give a system of {\it global} flat
coordinates on the moduli space of corresponding curves.
In \cite{D3} using the Hamiltonian approach to the Whitham
theory the integrability of general Witten-Dijgraagh-Verlinder-Verlinder (WDVV)

Two- and three-points correlation functions
\begin{equation}
\langle \phi_{\alpha} \phi_{\beta} \rangle=\eta_{\alpha \beta},
\ \ c_{\alpha \beta \gamma}=
\langle \phi_{\alpha} \phi_{\beta} \phi_{\gamma} \rangle \label{e1}
\end{equation}
of any topological field theory with $N$ primary fields $\phi_1,\ldots,\phi_N$
define an associative algebra
\begin{equation}
\phi_{\alpha} \phi_{\beta}=c_{\alpha \beta}^{\gamma}\phi_{\gamma},
\ c_{\alpha \beta}^{\gamma}=c_{\alpha \beta \mu}\eta^{\gamma \mu},
\ \eta_{\alpha \mu}\eta^{\mu \beta}=\delta_{\alpha}^{\beta}.\label{e2}
\end{equation}
with a unit $\phi_1$
\begin{equation}
\eta_{\alpha \beta}=c_{1 \alpha \beta }.\label{e2'}
\end{equation}
It turns out that there exists $N$ parametric deformation of the theory such
that ``metric" $\eta_{\alpha \beta}$ is a constant and three-point
correlators are given by the derivatives of free energy $F(t)$ of the deformed
theory
\begin{equation}
c_{\alpha \beta \gamma}(t)=\partial_{\alpha \beta \gamma}F(t),\ \ \
\eta_{\alpha \beta}=\partial_{1 \alpha \beta }F(t)=const.\label{e3}
\end{equation}

The associativity conditions of algebra (\ref{e2}) with structure constants
(\ref{e3}) are equivalent to a system of partial differential equations on $F$
(WDVV equations).
In \cite{D3} ``spectral transform" for these equations were proposed. It
proves their integrability, however (as it seems for us) the explicit
representation of all corresponding models remains an open problem.

In section 5 we show that each ``algebraic" orbit of the universal
Whitham hierarchy gives an exact solutions
of WDVV equations. Moreover, a generalization of W-constraints for
corresponding $\tau$-functions, that are proved in section 4, provide
some evidence that the universal Whitham hierarchy can be considered as
a universal (at tree-level) topological field theory coupled with gravity.

In this introduction we present a definition of the Whitham hierarchy
in a most general form. All ''integrable"  partial differential equations,
that are considered in the framework of the ''soliton"  theory, are equivalent
to compatibility conditions of auxiliary linear problems. The general
algebraic-geometrical construction of their exact periodic and quasi-periodic
solutions was proposed in \cite{K3}. There the concept of the Baker-Akhiezer
functions were introduced. (The analytical properties of
the Baker-Akhiezer functions are the generalization of
properties of the Bloch solutions of the finite-gap Sturm-Liouville operators,
which were found in a serious of papers by Novikov, Dubrovin, Matveev and
Its \cite{NDMI}).

The ``universal" set of algebraic-geometrical data is as follows. Consider the
space $\hat M_{g,N}$ of smooth algebraic curves $\Gamma_g$ of genus $g$ with
local coordinates $k_{\alpha}^{-1}(P)$ in neighbourhoods of
$N$ punctures $P_{\alpha}$ , ($k_{\alpha}^{-1}(P_{\alpha}) = 0$)
\begin{equation}
\hat M_{g,N} = \{\Gamma_g,P_{\alpha},k_{\alpha}^{-1}(P),\ \alpha=1,\ldots,N\}.
\label{(1.1)}
\end{equation}
This space is a natural bundle over the moduli space $M_{g,N}$ of smooth
algebraic curves $\Gamma_g$ of genus $g$ with $N$ punctures
\begin{equation}
\hat M_{g,N} = \{\Gamma_g,P_{\alpha},k_{\alpha}^{-1}(P)\}
\longmapsto  M_{g,N} = \{\Gamma_g,P_{\alpha}\}.\label{(1.2)}
\end{equation}
For each set of data (\ref{(1.1)}) and each set
of $g$ points $\gamma_1\ldots,\gamma_g$ on $\Gamma_g$ in a general position
( or, equivalently, for a point of the Jacobian $J(\Gamma_g)$ )
the algebraic-geometrical construction gives a quasi-periodic
solution of some integrable PNDE. (For given non-linear integrable equation the
corresponding set of data have to be specified. For example, the solutions
of the Kadomtsev-Petviashvili (KP) hierarchy corresponds to the case $N=1$. The
solutions of the two-dimensional Toda lattice corresponds to the case $N=2$.)

The data (\ref{(1.1)}) are ``integrals" of the infinite
``hierarchy" of integrable non-linear differential equations, that can
be represented as a set of commuting ``flows" on a phase space. Let $t_A$
be a set of all corresponding ``times". In the framework of the ``finite-gap"
(algebraic-geometrical) theory of integrable equations each time $t_A$ is
coupled with a meromorphic differential $d\Omega_A (P|{\cal M}),\ {\cal M}\in
\hat M_{g,N} $
\begin{equation}
t_A \longmapsto d\Omega_A (P|{\cal M})\label{no}
\end{equation}
that is ``responsible" for the flow. ( $ d\Omega_A (P|{\cal M})$ is
a differential with respect to the variable $P\in \Gamma$ depending on
the data (\ref{(1.1)}) as on external parameters.)

In \cite{K4} the algebraic-geometrical perturbation theory for integrable
non-linear (soliton) equations was developed. It was stimulated by the
application of the Whitham approach for (1+1) integrable equation of the
KdV type \cite{WFMD}. As usual in the perturbation theory, ''integrals" of an
initial equation become functions of the ``slow" variables
$\varepsilon t_A$ ($\varepsilon$ is a small parameter). ``The Whitham
equations" is a name for  equations which describe ``slow" variation
of ``adiabatic integrals". ( We would like to emphasize that
algebraic-geometrical approach represents only one side of the Whitham
theory. In \cite{ND} a deep differential-geometrical structure that is
assosiated with the Whitham equations were developed.)

Let $\Omega_A (k,T)$ be a set of holomorphic functions of the variable $k$
(which is defined in some complex domain $D$), depending on a finite or
infinite number of variables $t_A,\ T=\{t_A\}$. (We preserve the same
notation $t_A$ for slow variables $\varepsilon t_A$ because  we are
not going to consider ``fast" variables in this paper.) Let us introduce on the
space with coordinates $(k,t_A)$
a one-form
\begin{equation}
\omega=\sum_A \Omega_A(k,T)dt_A.\label{(1.3)}
\end{equation}
Its full external derivative equals
\begin{equation}
\delta \omega=\sum_A \delta \Omega_A (k,T)\wedge dt_A,\label{(1.4)}
\end{equation}
where
\begin{equation}
\delta \Omega_A=\partial_k \Omega_A dk+\sum_B
\partial_B \Omega_A dt_B,\ \partial_k=\partial/\partial k,\
\partial_A=\partial/\partial t_A. \label{(1.5)}
\end{equation}
The following equation
\begin{equation}
\delta \omega \wedge \delta \omega =0\label{(1.6)}
\end{equation}
we shall call by definition {\it the Whitham hierarchy}.

The ``algebraic" form (\ref{(1.6)}) of the Whitham equations is equivalent to a
 $A,B,C$
\begin{equation}
\sum_{\{A,B,C\}}\varepsilon^{\{A,B,C\}}\partial_A\Omega_B\partial_k
\Omega_C=0\label{(1.7)}
\end{equation}
(summation in (\ref{(1.7)}) is taken over all permutations of indecies $A,B,C$
a
$\varepsilon^{\{A,B,C\}}$ is a sign of permutation).

The equations (\ref{(1.6)}) are invariant with respect to an invertable change
of variable
\begin{equation}
k=k(p,T),\ \partial_p k\neq 0.\label{(1.8)}
\end{equation}
Let us fix some index $A_0$ and denote the corresponding function by
\begin{equation}
p(k,T)=\Omega_{A_0}(k,T).\label{(1.9)}
\end{equation}
At the same time we introduce the special notation for the corresponding
``time"
\begin{equation}
t_{A_0}=x.\label{(1.10)}
\end{equation}
After that all $\Omega_A$ can be cosidered as the functions of new
variable $p$,  $\Omega_A=\Omega_A(p,T)$. The equations (\ref{(1.7)}) for
$A,B,C=A_0$ have the form
\begin{equation}
\partial_A \Omega_B-\partial_B \Omega_A+\{\Omega_A,\Omega_B\}=0,\label{(1.11)}
\end{equation}
where $\{f,g\}$ denotes the usual Poisson bracket on the space of functions
of two variables $x,p$
\begin{equation}
\{f,g\}=\partial_x f\partial_p g-\partial_x g\partial_p f.\label{(1.12)}
\end{equation}

The Whitham equations were obtained in \cite{K4} in the form (\ref{(1.7)}).
In \cite{K5} it was noticed that they can be represented in the algebraic form
(\ref{(1.6)}). (We would like to mention here the papers \cite{tak}
where it was shown that algebraic form of the Whitham equations leads directly
to semiclassical limit of ``strings" equations.)

The Whitham equations in the form (\ref{(1.7)}) are equations on the set of
func
real content. It's necessary to show that they do define correct systems of
equations on the spaces $\hat M_{g,N} $. For zero-genus case
($g=0$) it will be done in the next section.
In the same section a construction (\cite{K4})of exact solutions of the
zero-genus
hierarchy corresponding to its ``algebraic" orbits are presented. The key
element of the scheme \cite{K4} is a construction of a potential $S(p,T)$ and
a ``connection" $E(p,T)$ such that after change of variable
\beq
p=p(E,T),\ \Omega_A(E,T)=\Omega_A(p(E,T),T)\label{(1.14)}
\eeq
the following equalities
\beq
\Omega_A(E,T)=\partial_A S(E,T).\label{(1.15)}
\eeq
are valid.

In section 3 the $\tau$-function for the Whitham equations on the spaces
$\hat M_{0,N}$  is introduced. For all genera (the case $g>0$ is considered in
section 7) $\tau$-function can be represented in the following ``field
theory" form
\beq
\tau =\int_{\Gamma}\bar d S\wedge dS.\label{(1.16)}
\eeq

{\it Important remark.} The integral (\ref{(1.16)}) does not equal to zero,
because $S(p,T)$ is holomorphic
on $\Gamma$ except at the punctures $P_{\alpha}$ and some contours, where
it has ``jumps". Therefore, the integral over $\Gamma$ equals to
a sum of the residues at $P_{\alpha}$  and the contour
integrals of the corresponding one-form.

The $\tau$-function is a function of the variables $t_A$, $\tau=\tau(T)$. As it
will be shown in the section 3 it contains a full information about the
corresponding solutions $\Omega_A(p,T)$ of the Whitham equations. For
$g>0$ in $\tau$ a part of geometry of moduli spaces is incoded.

In section 4 zero-genus Virasoro and W-constraints for $\tau$-function are
proved. In section 5 the primary chiral rings corresponding to algebraic orbits
of Whitham hierarchy are considered. The last section is preceded by section 6
where using ideas of \cite{tak} a ``direct transform" for general Whitham
hierarchy is discussed. It turns out that the existence of a potential $S$ is
not a characteristic property of the construction of solutions. In a hidden
form

All results that are proved for genus-zero Whitham hierarchy in the first
five sections are generalized for arbitrary genus case in section 7. We present
same way as in genus zero case but requier greater lenght due to pure technical

\section{Whitham hierarchy. Zero genus case}

In zero genus case a point of the ``phase space" $\hat M_{g=0,N}$ is a set
of points $p_{\alpha},\  \alpha=1,\ldots,N,$ and a set of formal local
coordinates $k_{\alpha}^{-1}(p)$
\beq
k_{\alpha}(p)=\sum_{s=-1}^{\infty}v_{\alpha ,s}(p-p_{\alpha})^s \label{(2.1)}
\eeq
(``formal local coordinate" means that r.h.s of (\ref{(2.1)}) is considered
as a formal series without any assumption on its convergency).
Hence, $\hat M_{0,N}$ is a set of sequences
\beq
\hat M_{0,N}=\{p_{\alpha},\ v_{\alpha ,s},\ \alpha=1,\ldots,N,\
s=-1,0,1,2,\ldots\}\label{(2.2)}
\eeq
The Whitham equations define a dependence of points of $\hat M_{0,N}$
with respect to the variables $t_A$ where the set of indecies $\cal A$
is as follows
\beq
{\cal A}=\{A=(\alpha ,i), \ \alpha=1,\ldots,N,\ i=1,2,\ldots\
{\rm and\ for}\ i=0,\alpha\neq 1\}.\label{(2.3)}
\eeq
As it was explained in the introduction we can fix one of the points
$p_{\alpha}$ with the help of a appropriate change of the variable $p$.
Let us choose: $p_1=\infty$.

Introduce meromorphic functions $\Omega_{\alpha,i}(p)$ for $i>0$
with the help of the following conditions:

$\Omega_{\alpha,i}(p)$ has a pole only at $p_{\alpha}$ and coincides
with the singular part of an expansion of $k_{\alpha}^i(p)$ near this
point, i.e.
$$
\Omega_{\alpha,i}(p)=\sum_{s=1}^i w_{\alpha,i,s}(p-p_{\alpha})^{-s}=
k_{\alpha}^i(p)+O(1),
$$
\beq
\Omega_{\alpha,i}(\infty)=0,\ \alpha\neq 1.\label{(2.4)}
\eeq
\beq
\Omega_{1,i}(p)=\sum_{s=1}^i w_{1,i,s}p^s=
k_{1}^i(p)+O(k_1^{-1}).\label{(2.5)}
\eeq
These polynomials can be written in the form of the Cauchy integrals
\beq
\Omega_{\alpha,i}(p,T)=\frac{1}{2\pi i}\oint_{C_{\alpha}}
\frac{k_{\alpha}^i(z_{\alpha},T)dz_{\alpha}}{p-z_{\alpha}}.\label{2.5'}
\eeq
Here $C_{\alpha}$ is a small cycle around the point $p_{\alpha}$.

The functions $\Omega_{\alpha,i=0}(p),\ \alpha\neq 1$ are equal to
\beq
\Omega_{\alpha,0}(p)=-\ln (p-p_{\alpha}).\label{(2.6)}
\eeq

{\it Remark.} The asymmetry of the definitions of $\Omega_{\alpha,i}$ reflects
our intention to choose the index $A_0=(1,1)$ as ``marked" index.

The coefficients of $\Omega_{\alpha,i>0}(p)$ are polynomial functions of
$ v_{\alpha ,s}$. Therefore, the Whitham equations (\ref{(1.5)}) (or
(\ref{(1.11)})) can be
rewritten as equations on $\hat M_{0,N}$. But still it have to be shown
that they can be considered as a correctly defined system.
\begin{th}
The zero-curvature form (\ref{(1.11)}) of the Whitham hierarchy in zero-genus
case is equivalent to the Sato-form that is a compatible system of evolution
 equations
\beq
\partial_A k_{\alpha}=\{k_{\alpha},\Omega_A\}.\label{1'}
\eeq
\end{th}
{\it Proof}. Consider the equations (\ref{(1.11)}) for $B=(\alpha,j>0)$.
 From the definition of $\Omega_A$ it follows that
\beq
\partial_A k_{\alpha}^j-\{k_{\alpha}^j,\Omega_A\}=
\partial_B\Omega_A-\{\Omega_A,\Omega_B^-\}+O(1).
\label{2'}
\eeq
Here and below we use the notation
\begin{eqnarray}
\Omega_A^-&=&\Omega_A-k_{\alpha}^i, \ \ \ \ \ \ {\rm for}\ \ A=(\alpha,i>0),
\nonumber \\
\Omega_{\alpha,0}^-&=&\Omega_{\alpha,0}-\ln k_{\alpha}. \label{3'}
\end{eqnarray}
Hence,
\beq
\partial_A k_{\alpha}^j-\{ k_{\alpha}^j,\Omega_A \} =0(k^{i-1}). \label{4'}
\eeq
Therefore,
\beq
\partial_A k_{\alpha}-\{k_{\alpha},\Omega_A\}=0(k^{i-j}).\label{5'}
\eeq
The limit of (\ref{5'}) for $j\to \infty$ proves (\ref{1'}).
The inverse statement that (\ref{1'}) is a correctly defined system can
be proved in a standart way. So we shall skip it.

Let us demonstrate a few examples.

\bigskip
\noindent{{\it Example 1. Khokhlov-Zabolotskaya hierarchy.}}
\bigskip

The Khokhlov-Zabolotskaya hierarchy is the particular $N=1$ case
of our considerations. Any local coordinate $K^{-1}(p)$ near infinity
($p_1=\infty$)
\beq
 K(p)=p+\sum_{s=1}^{\infty}v_sp^{-s} \label{(2.7)}
\eeq
defines a set of polynomials:
\beq
\Omega_i (p)\ =\ \bigl[K^i (p) \bigr]_+ , \label{(2.8)}
\eeq
here $[...]_+$  denotes a non-negative part of Laurent series.
For example,
\beq
\Omega_2 = k^2 + u,\ \ \Omega_3=k^3+{3\over2}uk+w,\ \  \
{\rm where}\ u=2v_1,\ \ w=3v_2.\label{(2.9)}
\eeq
If we denote $t_2=y,\ t_3=t$ then the equation (\ref{(1.11)})
for $A=2,\ B=3$ gives
\beq
w_x={3\over 4}u_y,\ \ w_y=u_t - {3\over 2} uu_x,\label{(2.10)}
\eeq
from which the dispersionless KP (dKP) equation (which is also called by the
Khokhlov-Zabolotskaya equation) follows :
\beq
 {3\over 4}  u_{yy} +\bigl( u_t - {3\over 2} uu_x \bigr)_x=0 .\label{(2.11)}
\eeq

The Khokhlov-Zabolotskaya equation is a partial diffe\-ren\-tial
equation and though it has not pure evolution form, one can expect that its
solutions have to be uniquelly defined by their Cauchy data $u(x,y,t=0)$,
that is a function of two variables $x,y$. Up to now it is not clear
if this two-dimensional equation can be considered as  the third equivalent
form of the Whitham hierarchy (we remind that formally solutions of
the hierarchy (\ref{1'}) depend on a infinite number of functions of one
variable).

\bigskip
\noindent{{\it Example 2. Longwave limit of 2d Toda Lattice.}}
\bigskip

The hierarchy of longwave limit of two-dimensional Toda equation are the
particular $N=2$ case of our considerations. There are two local parameters.
One
of them is near the infinity $p_1=\infty$ and one is near a point $p_2=a$.
They depend on two set of the variables $t_{\alpha,s},\ \alpha=1,2,
\ s=1,2,\ldots$ and also on the variable $t_0$. We shall present here
only the basic two-dimentional equation of this hierarchy (an analogue of
Khokhlov-Zabolotskaya equation).

Consider three variables $t=t_0,x=t_{1,1},y=t_{2,1}$. The corresponding
functions are
\beq
\Omega_0=\ln (p-a),\ \ \Omega_{1,1}=p,\ \
\Omega_{2,1}={v\over p-a}.\label{(t)}
\eeq
Their substitution into the zero-curvature equation (\ref{(1.11)}) gives
\beq
v_x=a_tv,\ \ v_t+a_y=0,\ \ w_t=0.\label{t2}
\eeq
 From (\ref{t2}) it follows
\beq
\partial^2_{xy}\phi+\partial^2_te^{\phi}=0, \ {\rm where}\ \
\phi=\ln v.\label{t3}
\eeq
This is a longwave limit of the 2d Toda lattice equation
\beq
\partial^2_{xy}\varphi_n=e^{\varphi_{n-1}-\varphi_n}
-e^{\varphi_n-\varphi_{n+1}}\label{t4}
\eeq
corresponding to the solutions that are slow functions of the discrete
variable $n$, which is replaced by the continuous variable $t$. The equation
(\ref{t3}) has arised independently in general relativity, in the theory of
wave
can be found in \cite{sav} where a representation of solutions of (\ref{t3})
in terms of convergent series were proposed.

\bigskip
\noindent{\it Example 3. N-layer solutions of the Benny equation.}
\bigskip

This example corresponds to a general $N+1$ points case, but we consider
only one zero-curvature equation. Let us choose three  functions
\beq
\Omega_1=p,\ \ \Omega_2=p+\sum_1^N {v_i\over p-p_i},\ \
\Omega_3=p^2+u,\label{b}
\eeq
which are coupled with the variables $x,y,t$, respectively. ( In our standart
notations they are
\beq
x=t_{1,1},\ \ y=\sum_{\alpha=1}^{N+1} t_{\alpha,1},\ \ t=t_{1,2}.)\label{b1}
\eeq
The zero-curvature equation (\ref{(1.11)}) gives the system
\beq
p_{it}-(p_i^2)_x+u_x=0,v_{it}=2(v_ip_i)_x,u_y-u_x+2\sum_iv_{ix}=0.\label{b2}
\eeq
Solutions of this system that do not depend on $y$ are
N-layer solutions of the Benny equation. As it was noticed in \cite{Z} the
corresponding system
\beq
p_{it}-(p_i^2)_x+u_x=0,\ \ v_{it}=2(v_ip_i)_x,\ \ u=2\sum_iv_i\label{b3}
\eeq
is a classical limit of the vector non-linear Schr\"odinger equation
\beq
i\psi_{it}=\psi_{i,xx}+u\psi_i,\ \ \ u=\sum_i|\psi_i|^2.\label{b4}
\eeq
(Using this observation in \cite{Z} the integrals of (\ref{b3}) were
found.)

\bigskip

In the second part of this section we consider ``algebraic"
orbits of the genus-zero Whitham equations. By definition they are specified
with the help of the constraint:
{\it there exists a meromorphic solution $E(p,T)$ of the equations
\beq
\partial_A E=\{E,\Omega_A\},\label{(1.13)}
\eeq
such that}
\beq
\{E(p,T),k_{\alpha}(p,T)\}=0.\label{25}
\eeq
The last equality implies that there exist functions $f_{\alpha}(E)$ of one
variable such that
\beq
k_{\alpha}(p,T)=f_{\alpha}(E(p,T)).\label{26}
\eeq

In order not to be lost in too general setting right at the begining, let us
beging with an example.

\bigskip
\noindent{\it Example. Lax reductions ( N=1)}
\bigskip

Consider solutions of the dKP hierarchy such that some power of
local parameter (\ref{(2.7)}) is a polynomial in $p$, i.e.
\beq
E(p,T) = p^n + u_{n-2} p^{n-2} + ... + u_0 = k_1^n(p,T). \label{(5.6)}
\eeq
The relation (\ref{(5.6)}) implies that only a few first coefficients of the
local parameter are independent. All of them are polynomials with respect to
the cofficients $u_i$ of the polynomial $E(p,T)$. The corresponding
solutions of dKP hierarchy can be described in terms of dispersionless
Lax equations
\beq
\partial_i E(p,T)=\{E(p,T),\Omega_i(p,T)\},\label{27}
\eeq
where
\beq
\Omega_i(p,T)=[E^{i/n}(p,T)]_+\label{29}
\eeq
(as before, $[\ldots]_+ $ denotes a non-negative part of corresponding Laurent
series).
These solutions of KP hierarchy can be also characterized by the property
that they do not depend on the variables $t_n,t_{2n},t_{3n},\ldots$. We are
going to construct an analoque of the ``hodograph" transform for
the solution of this equations. It is a generalization and effectivization
of a scheme \cite{Tzar}, where ``hodograph-type" transform were proposed
for hydrodinamic-type diagonalizable Hamiltonian systems (\cite{ND}).

Let us introduce a generating function
\beq
S(p)=\sum_{i=1}^\infty t_i \Omega_i (p)
=\sum_{i=1}^\infty t_i K^i + O(K^{-1}),\label{(5.11)}
\eeq
where $\Omega_i$ are given by (\ref{(2.8)}) and $K=E^{1/n}$. (If there is
only a finite number of $t_i$ that are not equal to zero, then $S(p)$ is a
polynomial.) The coefficients of $S$ are linear functions of $t_i$ and
polynomials in $u_i$. We introduce a dependence of $u_j$ on the variables
$t_i$ with the help of the following algebraic equations:
\beq
{dS\over dp}(q_s)=0,\label{(5.14)}
\eeq
where $q_s$ are zeros of the polynomial
\beq
{dE\over dp} (q_s)\ =\ 0.\label{(5.13)}
\eeq

{\it Remark.} It is not actually necessary to solve the equation (\ref{(5.13)})
in order to find $q_s$. We can choose $q_s,\ s=1,\ldots,n-1$ and $u_0$ as a new
set of unknown functions, due to the equality
\beq
{dE\over dp}=np^{n-1}+(n-2)u_{n-2}p^{n-3}+\ldots+u_1=\prod_{s=1}^n
(p-q_s),  \label{5.13`}
\eeq
$$q_n=-\sum_{s=1}^{n-1}q_s.$$
Let us prove that if the dependence of $E=E(p,T)$ with respect to the variables
 $t_i$ is defined from (\ref{(5.14)}), then
\beq
\partial_i S(E,T) = \Omega_i (E,T). \label{(5.15)}
\eeq
Consider the function $\partial_i S(E,T)$. From (\ref{(5.11)}) it follows that
\beq
\partial_i S(E)=K^i+O(K^{-1})=\Omega_i(E)+O(K^{-1}).\label{(5.16)}
\eeq
Hence, it is enough to prove that $\partial_i S(E)$ is a polynomial in $p$,
because by definition $\Omega_i$ is the only polynomial in $p$ such that
\beq
\Omega_i (p)=K^i+O(K^{-1}).\label{(5.17)}
\eeq
The function $\partial_i S(E,T)$ is holomorphic everywhere except may be at
$q_s(T)$. In a neighbourhood of $q_s(T)$ a local coordinate is
\beq
(E-E_s(T))^{1/2},\ \ \ E_s(T)=E(q_s(T))\label{5.17'}
\eeq
(if $q_s$ is a simple root of (\ref{(5.13)})).
Hence, a priori $S$ has the expansion
\beq
S(E,T)=\alpha_s(T)+\beta_s(T) (E-E_s(T))^{1/2}+\ldots \label{(5.18)}
\eeq
and the derivative $\partial_i S(E,T)$ might be singular at the points $q_s$.
But the defining relations (\ref{(5.14)}) imply that $\beta_s=0$. Therefore,
$\partial_i S(E)$ is regular everywhere except at the infinity and, hence, is
a polynomial. The equations (\ref{(5.15)}) are proved.

Let us present this scheme in another form. For each polynomial $E(p)$ of the
form (\ref{(5.6)}) and each formal series
\beq
Q(p)=\sum_{j=1}^{\infty} b_j p^j\label{q}
\eeq
the formula
\beq
t_i=\frac{1}{i}{\rm res}_{\infty}(K^{-i}(p)Q(p)dE(p)).\label{q1}
\eeq
defines the variables
\beq
t_k=t_k(u_i,b_j),\ \ i=0,\ldots,n-2,\ \ j=0,\ldots. \label{q2}
\eeq
as functions of the coefficients of $E,Q$.
Consider the inverse functions
\beq
u_i=u_i(t_1,\dots), \ b_j=b_j(t_1,\dots).\label{q3}
\eeq
{\it Remark.}
In order to be more precise let us consider a case when $Q$ is a polynomial,
i.e. $b_j=0,\ \ j>m$. From (\ref{q1}) it follows that $t_k=0,\ \ k>n+m-1$.
Therefore, we have $n+m-1$ ``times" $t_k,\ k=1,\ldots,n+m-1$ that are
linear functions of $b_j,\ j=1,\ldots,m$ and polynomials in $u_i,\
i=0,\ldots,n-2$. So, locally the inverse functions (\ref{q3}) are well-defined.
\begin{th}
The functions $u_i(T)$ are solutions of dispersionless Lax equation (\ref{27}).
Any other solutions of (\ref{27}) are obtained from this
particular one with the help of translations, i.e. $\tilde
u(t_i)=u(t_i-t_i^0)$.
\end{th}

\bigskip

Consider now general $N$ case. Let $E(p)$ be a meromorphic functions with
a pole of order $n$ at infinity and with poles of orders $n_{\alpha}$ at points
negative $n_{\alpha}\neq 0$, which means that $E$ has a zero of the order
 $-n_{\alpha}$, can be considered as well, but we are not going to do it here
in
\beq
E=p^n + u_{n-2} p^{n-2} + ... + u_0 +\sum_{\alpha=2}^M
\sum_{s=1}^{n_{\alpha}} v_{\alpha,s}(p-p_{\alpha})^{-s}.\label{d1}
\eeq
Consider a linear space of such functions, i.e. the space of sets
\beq
{\cal N}(n_{\alpha})=\{u_i, i=0,\ldots,n-2;\ \
v_{\alpha,s}, s=1,\ldots, n_{\alpha})\},\label{d2}
\eeq
$$\alpha=1,\ldots,M,\ \ n_1=n$$
If $N\leq M$  the function $E(p)$ defines the
local coordinates at the points $p_{\alpha}$ with the help of formula
\beq
k_{\alpha}^{n_{\alpha}}(p)=E(p), \alpha=1,\ldots,N.\label{d3}
\eeq
Therefore, ${\cal N}(n_{\alpha})$ can be identified with a subspace of
$\hat M_{0,N}$
\beq
{\cal N}\subset \hat M_{0,N}.\label{d4}
\eeq

\begin{th}
The subspaces ${\cal N}_M(n,\ n_{\alpha},\alpha=1,\ldots,M)$ are
invariant with respect to the Whitham equations on $\hat M_{0,N}$
that coincide with the flows
\beq
\partial_A E(p,T)=\{E(p,T),\Omega_A(p,T)\}, \label{d5}
\eeq
where $k_{\alpha}$ and $\Omega_A$ are defined with the help of formulae
(\ref{d3},\ref{2.5'}), respectively.  General solutions of (\ref{d5}) are
given in an implicite form with the help of the following algebraic equations
\beq
{dS\over dp}(q_s,T)=0, \ \ {\rm where} \ \ S(p,T)=\sum_A
(t_A-t_A^0)\Omega_A(p,T
\eeq
 which have to be fulfilled for all zeros $q_s$ of the function
\beq
{dE\over dp}(q_s)=0.\label{d7}
\eeq
\end{th}
The proof is the same as the proof of the previous theorem. Its main step is
a proof that from the defining relations (\ref{d6}) it follows that
\beq
\partial_A S(E,T)=\Omega_A(E,T).\label{d7'}
\eeq

{\it Definition.} The particular solutions of the Whitham hierarchy
corresponding to the algebraic orbits (\ref{d5}) and for which $t_A^0=0$ will
be

An alternative formulation of this theorem can be done in the following form.
Let $Q(p)$ be a meromorphic function  with poles at the points
 $p_{\alpha}$, i.e.
\beq
Q(p)=\sum_{j=1}^{\infty}b_{1,j}p^j+
\sum_{\alpha=2}^M\sum_{j=1}^{\infty}b_{\alpha,j}(p-p_{\alpha})^{-j}.
\label{d8}
\eeq
The formulae
\begin{eqnarray}
t_{\alpha,i}&=&\frac{1}{i}{\rm res}_{\alpha}(k_{\alpha}^{-i}(p)Q(p)dE(p))),
\ \ \  i>0; \nonumber\\
t_{\alpha,0}&=&{\rm res}_{\alpha}(Q(p)dE(p))\label{d9}
\end{eqnarray}
defines ``times" $t_{\alpha,i}$ as functions of the coefficients of $Q(p)$ and
$E(p)$ (which has the form (\ref{d1})). Consider the inverse functions
\beq
v_{\alpha,s}=v_{\alpha,s}(t_{\beta,i}), \ \ b_{\alpha,j}= b_{\alpha,j}
(t_{\beta,i}).\label{d10}
\eeq
(we remind that index $\alpha=1$ corresponds to the infinity $p_1=\infty$).

\begin{cor}
The inverse functions $v_{\alpha,s}(t_{\beta,i})$ are solutions of Whitham
equations. In particular, (for all $N$)
\beq
u(x,y,t)=\frac{2}{n}u_{n-2}(t_{1,1}=x,t_{2,1}=y,t_{3,1}=t,\ldots)
\label{d11}
\eeq
is a solution of Khokhlov-Zabolotskay equation (\ref{(2.11)}) and
\beq
\phi(x,y,t)=\frac{1}{n_2}\ln v_{2,n_2}(x=t_{1,1},y=t_{2,1},t=t_{2,0},\ldots)
\label{d12}
\eeq
is a solution of longwave limit of $2d$ Toda lattice equation (\ref{t3}).
\end{cor}
(We would like to emphasize that that in the formulae (\ref{d11},\ref{d12}) all
``times" except of the first ones are parameters.)

Inverse functions (\ref{d10}) define the dependence on the variables $t_A$
of the functions $E(p,T)$ and $Q(p,T)$. The differential of the potential
$S(p,T)$ equals
\beq
dS(p,T)=Q(p,T)dE(p,T).\label{d13}
\eeq
 From (\ref{d7'}) it follows that
\beq
\partial_AQ(E,T)={d\Omega_A(E,T)\over dE}.\label{d14}
\eeq
In particular,
\beq
\partial_x Q(E,T)={dp(E,T)\over dE}. \label{d15}
\eeq
The derivatives with fixed $E$ and $p$ are related with each other with
 the help of the following relation
\beq
\partial_Af(p,T)=\partial_Af(E,T)+{df\over dE}\partial_AE(p,T).\label{d16}
\eeq
Using (\ref{d15}) and (\ref{d16}) we obtain
\begin{cor}
The functions $Q(p,T)$ and $E(p,T)$ satisfy the quasi-classical ``string
equation"
\beq
\{ Q,E \}=1.\label{st}
\eeq
\end{cor}
(This corollary was prompted by \cite{tak} and we shall return to it at
greater lenght in section 6.)

\section{$\tau$-function}

In the previous section the ``algebraic" solutions of the Whitham hierarchy
in zero-genus case were constructed. It was shown that for
their ``potentials"
\beq
S(p,T)=\sum_A (t_A-t_A^0)\Omega_A(p,T)\label{32}
\eeq
(where $t_A^0$ are corresponding constants) the following equalities
\beq
\Omega_B(k_{\alpha},T)=\partial_B S(k_{\alpha},T),\ \ B=(\beta,i)\label{33}
\eeq
are fulfilled. In this section we define with the help of $S(p,T)$ the
$\tau$-functions corresponding to algebraic solutions of the Whitham equations.

The $\tau$-function of the universal Whitham hierarchy (in zero genus case)
would be by definition
$$\log\tau(T)=F(T),$$
\beq
F={1\over 2}\sum_{\alpha=1}^N {\rm res}_{\alpha}
(\sum_{i=1}^{\infty} \tilde t_{\alpha,i} k_{\alpha}^idS(p,T))+
\tilde t_{\alpha,0}s_{\alpha}(T),\label{35}
\eeq
$$\tilde t_{\alpha,i}=t_{\alpha,i}-t_{\alpha,i}^0,$$
where ${\rm res}_{\alpha}$ denotes a residue at the point $p_{\alpha}$ and
$s_{\alpha}$ is the coefficient of the expansion
\beq
S(p,T)=\sum_{\alpha=1}^N\sum_{i=1}^{\infty} \tilde t_{\alpha,i}k_{\alpha}^i
+t_{\alpha,0} {\rm ln}k_{\alpha}+s_{\alpha}+O(k^{-1})).\label{34}
\eeq
Here and below we use the notaion
\beq
t_{1,0}=-\sum_{\alpha=2}^N t_{\alpha,0}.\label{34'}
\eeq
The $\tau$-function can be rewritten in more compact form. Let us make a cuts
connecting the point $p_1=\infty$ with the points $p_{\alpha}$. After that
we can choose a branch of the function $S(p,T)$. The coefficient $s_{\alpha}$
equals
\beq
s_{\alpha}=\frac{1}{2\pi i}\oint_{\sigma_{\alpha}} {\ln}(p-p_{\alpha})dS,
\label{351}
\eeq
where $\sigma_{\alpha}$ is a contour around the corresponding cut.
The function $S$  has jumps on the cuts.
Its $\bar {\partial}$-derivative is a sum of delta-functions and its
derivatives
at the points $p_{\alpha}$ and one-dimentional delta-functions on cuts.
Therefore, (\ref{35}) can be represented in such a form:
\beq
F=\int \bar d S\wedge dS.\label{36}
\eeq
(The integral in (\ref{36}) is taken over the hole complex plane of the
variable
$p$. It is non-zero because $S(p,T)$ is holomorphic outside punctures and cuts,

\begin{th}
For the above defined $\tau$-function the following
equalities are fulfilled:
\beq
\partial_{\alpha,i} F(T)={\rm res}_{\alpha}(k_{\alpha}^i dS(p,T)),
\ \ i>0,\label{37}
\eeq
\beq
\partial_{\alpha,0} F(T)=s_{\alpha}.\label{38}
\eeq
\end{th}
{\it Proof}. Let us consider the derivative for $A=(\alpha, i>0)$. It equals
\beq
2\partial_A F= {\rm res}_{\alpha} (k_{\alpha}^i dS)+
\sum_{\beta=1}^N \sum_{j=1}^{\infty}
{\rm res}_{\beta}(\tilde t_{\beta,i}k_{\beta}^j d\Omega_A)
+\tilde t_{\beta,0}\Omega_A(p_{\beta}).\label{39}
\eeq
We use in(\ref{39}) the equality
\beq
\partial_A s_{\beta}=\Omega_A (p_{\beta}). \label{40}
\eeq
 From
\beq
 \sum_{\alpha=1}^N {\rm res}_{\alpha}
(\Omega_A d\Omega _B)=0\label{41}
\eeq
it follows that
\beq
{\rm res}_{\beta}(k_{\beta}^j d\Omega_{\alpha,i})=
{\rm res}_{\alpha}(k_{\alpha}^i d\Omega_{\beta,j}), \ \ j>0. \label{42}
\eeq
Besides this,
\beq
\Omega_A(p_{\beta})={\rm res}_{\beta}(\Omega_A d\ln(p-p_{\beta}))=
{\rm res}_{\alpha}(k_{\alpha}^id\Omega_{\beta,0}).\label{43}
\eeq
The substitution of (\ref{42},\ref{43}) into (\ref{39}) proves (\ref{37}). The
proof of (\ref{38}) is absolutely analoqous.

The formulae (\ref{37},\ref{38}) show that an expansion of $S(p,T)$
at the point $p_{\alpha}$ has the form
\beq
S(p,T)=\sum_{\alpha=1}^N\sum_{i=1}^{\infty} \tilde t_{\alpha,i}k_{\alpha}^i
+\tilde t_{\alpha,0}{\rm ln}k_{\alpha}+
\partial_{\alpha,0}F+\sum_{j=1}^{\infty}\frac 1j \partial_{\alpha,j}F
k_{\alpha}^{-j}.\label{44}
\eeq
 From (\ref{37},\ref{38}) it follows:
\begin{cor}
The second derivatives of $F$ are equal to
\beq
\partial_{A,B}^2F(T)=res_{\alpha}(k_{\alpha}^i d\Omega_B),\ A=(\alpha,i>0),
\label{45}
\eeq
\beq
\partial_{\alpha,0}\partial_{\beta,0} F(T)={\rm ln}(p_{\alpha}-p_{\beta}).
\label{46}
\eeq
\end{cor}
Hence, an expansion of the non-positive part (\ref{3'})of $\Omega_A(p)$ at the
point
$p_{\beta}$ has the form
\beq
\Omega_A^-=\partial_{\beta,0}F+\sum_{j=0}^{\infty}\frac 1j
(\partial_A \partial_{\beta,j}F) k_{\beta}^j. \label{47}
\eeq
Therefore, the $\tau$-function that depends on ``times", only, contains a
complete information on the functions $\Omega_A$.

\section{Truncated Virasoro and W-constraints}

In section 3 it was shown that any solution of the Whitham equations ($g=0$)
corresponding to an algebraic orbit can be obtained from ``homogeneous"
solution with the help of translations $\tilde t_A=t_A-t_A^0$. In this
section we consider $\tau$-functions of homogeneous solutions, only.

The truncated Virasoro constraints for the $\tau$-function of the
dispersionless
an invariance of residues with  respect to a change of variables. The same
approach can be applied for the general $N$-case, also. In this paper we use
another way that was inspired by the $N\to \infty$ limit of the loop-equations
for the one-matrix model ( a review of recent developments of the
loop-equations technique can be found in \cite{mak}).

The function $E(p)$ of the form (\ref{d1}) represents the complex plane of
the variable $p$ as D-sheet branching covering of the complex plane of
the variable $E$, $D=\sum_{\alpha}n_{\alpha}$. The zeros $q_s$ of $dE(q_s)=0$
are branching points of the covering. Hence, any function $f(p)$ can be
consider
are zeros of the equation
\beq
E(p_i)=E. \label{v1}
\eeq
The symmetric combination of the values $f(p_i)$
\beq
\tilde f(E)=\sum_{i=1}^D f(p_i) \label{v1'}
\eeq
is a single-valued function of $E$. Let us apply this argument for the function
$Q^K(p)$, where
\beq
Q(p)=\frac{dS(p)}{dE(p)}; \label{v2}
\eeq
$S(p)$ is the potential of the homogeneous solution of the Whitham
equations. The defining algebraic relations (\ref{d6}) imply that $Q(p)$ is
holomorphic outside the punctures $p_{\alpha}$ (that are ``preimages" of
the infinity $E(p_{\alpha})=\infty$). Therefore, the function
\beq
\tilde Q^K(E)=\sum_{i=1}^N Q^K(p_i), \label{v3}
\eeq
is an enteir function of the variable $E$. In other words the Laurent expansion
\beq
{\rm res}_{\infty} (\sum_{i=1}^N Q^K(p_i) E^{m+1} dE )=0,
\ \ m=-1,0,1,\ldots \ .\label{v4}
\eeq
The residue (\ref{v4}) at the infinite of the complex plane of $E$ is equal
to a sum of the residues at the points $p_{\alpha}$, i.e.
\beq
\sum_{i=1}^N{\rm res}_{\alpha} ( ( Q^K(k_{\alpha})
 E^{m+1} dE )=0, \ \ E=k_{\alpha}^{n_{\alpha}}. \label{v5}
\eeq
 From (\ref{44}) it follows that the function $Q(p)$ has
the expansion
\beq
Q(k_{\alpha})=\frac{1}{n_{\alpha}} \sum_{i=1}^{\infty}
 it_{\alpha,i}k_{\alpha}^{i-n_{\alpha}}+
t_{\alpha,0}k_{\alpha}^{-n_{\alpha}}+\sum_{j=1}^{\infty}
\partial_{\alpha,j} F k_{\alpha}^{-j-n_{\alpha}}. \label{v6}
\eeq
at the point $p_{\alpha}$.
The substitution of (\ref{v6}) into (\ref{v5}) for $K=1$ gives obvious
identities:
$$
\sum_{\alpha=1}^N t_{\alpha,0}=0,\ \  m=-1,
$$
\beq
\sum_{\alpha=1}^N \partial_{\alpha,mn_{\alpha}} F =0,\ \ m=0,1,\ldots.
\label{v7}
\eeq
(For Lax reductions $N=1$  the equalities (\ref{v7}) imply that F
does not depend on $t_n,t_{2n},\ldots$.) For $K>1$ the relations (\ref{v5})
lead to highly nontrivial equations. For example, the case $K=2$ corresponds to
\begin{th}
The $\tau$-function of the homogeneous solution of the Whitham equations
(corresponding to the orbit ${\cal N}(n_{\alpha})$) is a solution of the
equations
\beq
\sum_{\alpha=1}^N \frac{1}{n_{\alpha}} \bigl(
\sum_{i=n_{\alpha}+1}^{\infty} it_{\alpha,i}
\partial_{\alpha,i-n_{\alpha}} F +n_{\alpha}t_{\alpha,0}t_{\alpha,n_{\alpha}}+
{1\over 2}
\sum_{j=1}^{n_{\alpha}-1} j(n_{\alpha}-j)t_{\alpha,j}t_{\alpha,n_{\alpha}-j}
\bigr)=0; \label{v8}
\eeq
\beq
\sum_{\alpha=1}^N \frac{1}{n_{\alpha}} \sum_{i=1}^{\infty}it_{\alpha,i}
\partial_{\alpha,i} F+\frac12 t_{\alpha,0}^2=0;
\label{v9}
\eeq
\beq
\sum_{\alpha=1}^N \frac{1}{n_{\alpha}} \bigl(
\sum_{i=1}^{\infty}it_{\alpha,i}
\partial_{\alpha,i+mn_{\alpha}} F+\frac12 \sum_{j=1}^{mn_{\alpha}-1}
\partial_{\alpha,j} F\partial_{\alpha,mn_{\alpha}-j} F \bigr)
=0,\ \ m=1,\ldots.
\label{v10}
\eeq
\end{th}
The equalities (\ref{v4}) for any $K$ can be written in the form
\beq
 \sum_{\alpha=1}^N n_{\alpha}^{1-K}
\sum_{I,J} [i_1]t_{\alpha,i_1}\cdots [i_s]t_{\alpha,i_s}
\partial_{\alpha,j_{s+1}}F\cdots \partial_{\alpha,j_{K}}F=0, \label{v11}
\eeq
where the second sum is taken over all sets of indecies
$I=\{i_k\},\ J=\{j_k\}$ such that
\beq
\sum_{k=1}^s i_k=\sum_{k=s+1}^K j_k-(m+K-2)n_{\alpha},\ \ m>-1.\label{v12}
\eeq
and $[i]$ denotes
\beq
[i]=i,\ \ {\rm if}\ i\neq 0;\ \ [0]=1. \label{v13}
\eeq
For $N=1$ the equations (\ref{v11}) coincide with a nonlinear part of
the $W_K$ constraints.

At the end of this section we present the truncated Virasoro constraints for
$N=1$ and $n=2$ in the form of the planar limit of the loop-equations
for one-matrix hermitian model.

Consider the negative part of $Q(k)$ ($N=1,n=2$)
\beq
-{\cal W}_0=Q^-(k)=\frac{1}{2} \sum_{i=1}^{\infty}
 t_1 k^{-1}+\sum_{j=1}^{\infty}
\partial_{2j-1} F k^{-2j-1}\label{v14}
\eeq
and introduce
\beq
V(k)=\sum_{i=1}^{\infty} \tilde t_{2i}k^{2i}, \label{v15}
\eeq
where
\beq
\tilde t_{2i}=\frac{2i+1}{2i} t_{2i+1}.\label{v16}
\eeq
Then
\beq
Q(k)=V'(k)-{\cal W}_0 \label{v17}
\eeq
and (\ref{v5}) is equivalent to the equation
\beq
\oint_C {V'(\xi ){\cal W}_0(\xi)\over k-\xi}d\xi=
\frac12 {\cal W}_0^2, \label{v18}
\eeq
where $C$ is a small contour around the infinity. (\ref{v18}) is a planar limit
(see \cite{mak})
\beq
\oint_C {V'(\xi ){\cal W}(\xi)\over k-\xi}d\xi=
\frac12 {\cal W}^2 +\sum_{i=1}^{\infty}k^{-2i-1}{\partial {\cal W}\over
\partial t_{2i}}. \label{v19}
\eeq
In (\ref{v19}) ${\cal W}(k)$ is the Wilson loop-correlator that by
definition is equal to
\beq
{\cal W}=\langle {\rm tr}\frac 1{k-X} \rangle =
\int {\rm tr}\frac 1{k-X} e^{-{\rm tr}V(X)}dX, \label{v20}
\eeq
where $X$ is a hermitian $M\times M$ matrix.

{\it Remark.} As it was shown in \cite{dug} the double-scaling limit of $n-1$
matrix chain model is related with the $n$-th reduction of the KP-equation.
Dispersionless Lax equations (\ref{(5.6)},\ref{27}) are their classical limit.
Therefore, the negative part ${\cal W}_0$ of the series (\ref{v6}) for
$N=1$ and arbitrary $n$ has to be related with the planar limit of some
Wilson-type correlators for multi-matrix models. Therefore,
higher ``loop-equations" (corresponding to $K>2$) have to be fulfilled for
them. It should be interesting to find a direct way to produce the
corresponding equations in the framework of the multimatrix models.

\section{Primary rings of the topological field theories}

Topological minimal models were introduced in \cite{Eg} and were considered in
\cite{Li}. They are a twisted version
of the discrete series of $N=2$ superconformal Landau-Ginzburg (LG) models. A
large class of the $N=2$ superconformal LG models has been studied in
\cite{vafa}.
It was shown, that a finite number of states are topological, which means that
their operator products have no singularities. These states form a closed ring
$\cal  R$, which is called a primary chiral ring. It can be expressed in terms
of the superpotential $E(p_i)$ of the corresponding model
\beq
 {\cal R} = {C[p_i]\over dE=0} ,\ \ \
dW= {\partial E\over \partial p_i } dp_i .\label{6.1}
\eeq
In topological models these primary states are the only local physical
excitations.

In \cite{VVD}, it was shown that correlation functions of primary chiral fields
can be expressed in terms of perturbed superpotentials $E(p_i,t_1,t_2,...)$.
For $A_{n-1}$ model the unperturbed superpotential has the form:
\beq
E_0 =  p^n . \label{6.2}
\eeq
The coefficients of a perturbed potential
\beq
E(p)=p^n + u_{n-2} p^{n-2} +...+u_0  \label{6.3}
\eeq
can be considered as the coordinates on the space of deformed topological
minimal models. In \cite{VVD} the dependence of $u_i$ on the coordinates
$t_1,\ldots,t_{n-1}$ that are ``coupled" with primary fields $\phi_i$
were found. It was shown that the deformation of the ring ${\cal R}$
\beq
{\cal R}(t_1,\ldots,t_{n-1})=C[p]/(dE(p,t_1,\ldots,t_{n-1})=0) \label{T1}
\eeq is a potential deformation of the Fr\"obenius algebra ( in the sense
that was explained in Introduction).

In this section we consider an application of the general Whitham equations
on $\hat M_{0,N}$ to the theory of potential deformations of the Fr\"obenius
algebras. They are based on the following formula for the third
logariphmic derivatives of the $\tau$-function. Let $E(p,T)$ be a
homogeneous solution of the Whitham equations (\ref{d5}) corresponding to
an algebraic orbit ${\cal N}(n_{\alpha})$, i.e.  $E(p)$ has the form
(\ref{d1})
$$E=p^n + u_{n-2} p^{n-2} + ... + u_0 +\sum_{\alpha=2}^M
\sum_{s=1}^{n_{\alpha}} v_{\alpha,s}(p-p_{\alpha})^{-s}.$$
The formulae (\ref{d9},\ref{d10}) define the dependence $E(p)$ and
the ``dual" function $Q(p)$ with respect to the variables $t_A$.
\begin{th}
The third logariphmic derivatives of the $\tau$-function of the homogeneous
solution of the Whitham equation corresponding to an algebraic orbit
${\cal N}(n_{\alpha})$ are equal to
\beq
\partial_{ABC}^3 F=\sum_{q_s} {\rm res}_{q_s} \bigl(
{d\Omega_A d\Omega_B d\Omega_C \over dQ\  dE} \bigr), \label{r3}
\eeq
where $q_s$ are zeros of the differential $dE(q_s)=0$.
\end{th}
{\it Proof.} Let us suppose that $A=(\alpha,i>0)$. (The case when
$A,B,C$ are equal to $(\alpha,0),(\beta,0),(\gamma,0)$ can be considered in the
same way.) From (\ref{37}) it follows that
\beq
\partial_C \partial_{AB}^2F={\rm res}_{\alpha}(k_{\alpha}^id\partial_C
 \Omega_B)=-{\rm res}_{\alpha}(\partial_C\Omega_Bd\Omega_A).\label{r4}
\eeq
Here the derivative $\partial_C\Omega_A(E,T)$ is taken for the fixed $E$. As
it was explained in section 2 the function $E(p,T)$ is a ``good" coordinate
except at the points $q_s(T)$ where local coordinates have the form
(\ref{5.17'}). Hence, at the point $q_s(T)$ the function $\Omega_A$ has
the expansion
\beq
\Omega_B(E,T)=w_{B,0}(T)+w_{B,1}(T)(E-E_s(T))^{1/2}+\cdots, \label{r5}
\eeq
$$E_s(T)=E(q_s(T),T).$$
A sum of all residues of a meromorphic differential equals zero. Therefore,
\beq
\partial_{ABC}^3 F=\sum_{q_s} {\rm res}_{q_s}(\partial_C\Omega_Bd\Omega_A).
\label{r6}
\eeq
 From (\ref{r5}) we have that in a neighbourhood of the point $q_s$
\beq
\partial_C \Omega_B=-\partial_C E_s {d\Omega_B\over dE} +O(1).\label{r7}
\eeq
Therefore,
\beq
{\rm res}_{q_s}(\partial_C\Omega_Bd\Omega_A)=-
{\rm res}_{q_s} \bigl( \partial_C E_s {d\Omega_A d\Omega_B \over dE} \bigr).
\label{r8}
\eeq
 From (\ref{d5}) it follows
\beq
\partial_C E_s=\partial_x E_s {d\Omega_C\over dp}(q_s). \label{r9}
\eeq
The string equation (\ref{st}) implies
\beq
\partial_x E_s=-{dp\over dQ}(q_s).\label{r10}
\eeq
Substitution of (\ref{r9}) and (\ref{r10}) into (\ref{r8}) proves the theorem.

For each algebraic orbit ${\cal N}(n_{\alpha})$ let us define a ``small
phase" space (see motivation in \cite{W1}). It will be a space of times $t_a$
with the indecies $a$ belonging to a subset ${\cal A}_{sm}$
\beq
{\cal A}_{sm}=\{(\alpha,i)| \alpha=1, i=1,\ldots, n-1;
\alpha=2,\ldots,N,i=0,\ldots,n_{\alpha} \}. \label{T2}
\eeq
Let us fix all other times $t_A$
$$t_{1,n}=0,\ t_{1,n+1}=\frac{n}{n+1},\ t_{1,i}=0,\ i>n+1;$$
\beq
t_{\alpha,i}=0, \ \alpha=2,\ldots,N,\ i>n_{\alpha}.\label{T3}
\eeq
Comparision with (\ref{v6}) shows that in this case
\beq
Q(p)=p \label{T4}
\eeq
in the formula (\ref{d9}). In other words $t_a$ as a functions of
$u_i,\  v_{\alpha,s}$ are given by the formulae
\begin{eqnarray}
t_{\alpha,i}&=&\frac{1}{i}{\rm res}_{\alpha}(k_{\alpha}^{-i}(p)pdE(p))),
\ \ \  i>0; \nonumber\\
t_{\alpha,0}&=&{\rm res}_{\alpha}(pdE(p))\label{T5}
\end{eqnarray}
Inverse functions define a dependence of the coefficients of $E$ with respect
to the variables $t_a$
\begin{cor}
Let
\beq
F(t_a)=F(t_a,t_{1,n+1}={n\over n+1}, t_A=0,\ A\not\in{\cal A}_{sm})
\label{T6}
\eeq
be the restriction of $F=\ln \tau$ on the affine space that is
${\cal A}_{sm}$ shifting by $t_{1,n+1}={n\over n+1}$. Then
\beq
\partial_{abc}^3 F=\sum_{q_s} {\rm res}_{q_s} \bigl(
{d\Omega_a d\Omega_b d\Omega_c \over dp\  dE} \bigr).\label{T7}
\eeq
\end{cor}
Let us summarize the results. Each meromorphic function $E(p)$ of the form
(\ref{d1}) defines a factor ring
\beq
{\cal R}_E=\hat{\cal R}/(dE=0) \label{T8}
\eeq
of the ring $\hat{\cal R}$ of all meromorphic functions that are regular at
the zeros $q_s$ of the differential $dE$. The formula
\beq
\langle f,g\rangle=\sum_{q_s} {\rm res}_{q_s} \bigl( {f(p)g(p)\over E_p}dp
\bigr),\ f(p),\ g(p) \subset \hat{\cal R},\label{T9}
\eeq
defines a non-degenerate scalar product on ${\cal R}_E$. The scalar
product (\ref{T9}) supplies ${\cal R}_E$ by the structure of
the Fr\"obenius algebra. In the basis
\beq
\phi_a={d\Omega_a\over dp}\label{T10}
\eeq
the scalar product has the form
\beq
\langle \phi_a \phi_b \rangle=\eta_{ab}={[i][j]\over n_{\alpha}}
\delta_{\alpha,\beta} \delta_{i+j,n_{\alpha}}, \label{T11}
\eeq
where $[i]$ is the same as in (\ref{v13}). Our last statement is that the
formulae (\ref{T5}) define in an implicite form the potential deformations of
these Fr\"obenius algebras.

The case $N=1$ covers the results of \cite{VVD}. As it was mentioned in the
Introduction an integrability of WDVV equations was proved in \cite{D3}.
The results of this section can be considered as a explicit construction of
their particular solutions.

The process of a ``coupling" the ring ${\cal R}_E$ with topological
gravity corresponds to the process of ``switching on" of {\it all}
times of the Whitham hierarchy. It follows from the recurrent formula
for the third derivatives of $\tau$-function. First of all let us present
a formula
\beq
\Omega_{\alpha,i>0}dE=\frac{n_{\alpha}}{i+n_{\alpha}}
d\Omega_{\alpha,i+n_{\alpha}}+
\sum_{b=(\beta,j)\in {\cal A}_{sm}}
{n_{\beta}\over [n_{\beta}-j][j]} (\partial_{A,b}^2 F ) d\Omega_{\beta,
n_{\beta}-j}. \label{T12}
\eeq
It can be proved in a following way. The right and the left hand sides of
(\ref{T12}) are holomorphic ounside the punctures. Hence, it is enough
to compair their expansions at the points $p_{\alpha}$. The coefficients
of an expansion of $\Omega_A$ are given by the second derivatives of $F$
(\ref{47}). Therefore, (\ref{T12}) is fulfilled.

Let us denote for each $a=(\alpha,i>0)\in {\cal A}_{sm}$ the fields
\beq
{d\Omega_{\alpha,pn_{\alpha}+i}\over dQ} =\sigma_p(\phi_a).\label{T13}
\eeq
then the substitution of (\ref{T12}) into (\ref{r6}) proves the recurrent
formular for the correlation function for the gravitational decendants
\cite{W1}
\beq
\langle \sigma_p(\phi_a)\sigma_B \sigma_C \rangle=
\langle \sigma_{p-1}(\phi_a)\phi_b\rangle \eta^{bc}
\langle \phi_c \sigma_B \sigma_C\rangle, \label{T14}
\eeq
where $\sigma_B ,\sigma_C$ are any other states. ( The integrability of
general decendant equations were proved in \cite{D3}.)

{\it Remark.} This paper had been already written when author got a preprint
\cite{D1} where the Frobenious algebras and their ``small phase"
deformations corersponding to the Whitham hierarchy for the  multi-puncture
case

\section{Generating form of the Whitham equations}

In this short section (or rather a long remark) we would like to clarify
our construction of the algebraic solutions of the Whitham equations and
the definition of the corresponding $\tau$-function. It was stimulated by
the papers \cite{tak} where using our approach (\cite{K1}) the $\tau$-function
for longwave limit of 2d Toda lattice were introduced.

Let $\Omega_A (k,T)$ be a solution of the general zero-curvature equation
(\ref{(1.11)})
\beq
\partial_A \Omega_B-\partial_B \Omega_A+\{\Omega_A,\Omega_B\}=0,\label{g1}
\eeq
They are a compatibility condition for the equation
\beq
\partial_A E=\{E,\Omega_A\},\label{g2}
\eeq
Therefore, an arbitrary function $E(p,x)$ defines (at least locally) the
corresponding solution $E(p,T)$ of (\ref{g2}),
$E(p,x)=E(p,t_{A_0}=x,t_A=0,A\neq A_0)$.
In the domain  where $\partial_p E(p,T)\neq 0$ we can use
a variable $E$ as a new coordinate, $p=p(E,t)$. From (\ref{d16}) it follows
that in the new coordinate (\ref{g1}) are equivalent to the equations
\beq
\partial_A\Omega_B(E,T)=\partial_B \Omega_A(E,T).\label{g3}
\eeq
Hence, there exists a potential $S(E,T)$ such that
\beq
\Omega_A(E,T)=\partial_A S(E,T).\label{g4}
\eeq
Using this potential the one-form $\omega$ (\ref{(1.3)}) can reperesented as
\beq
\omega=\delta S(E,T)-Q(E,T)dE,\label{g5}
\eeq
where
\beq
Q(E,T)={\partial S(E,T)\over \partial E} .\label{g6}
\eeq
Hence,
\beq
\delta \omega=\delta E\wedge \delta Q.\label{g7}
\eeq
The formulae (\ref{d13}-\ref{d16}) that are valid in general case prove that
the functions $E$ and $Q$ as function of two variables $p,x$ satisfy the
classical string equation
\beq
\{Q,E\}=1.\label{g8}
\eeq
They show that
\beq
\partial_A Q=\{Q,\Omega_A\}.\label{g9}
\eeq

A set of pairs of functions $Q(p,x), E(p,x)$ satisfying the string equation is
a group with respect to the composition, i.e. if $Q(p,x), E(p,x)$ and
$Q_1(p,x), E_1(p,x)$ are solutions of (\ref{g8}) then the functions
\beq
\tilde Q(p,x)=Q_1(Q(p,x),E(p,x);\\ \tilde E(p,x)=E_1(Q(p,x),E(p,x)) \label{g9'}
\eeq
are solution of (\ref{g8}) as well. The Lie algebra of this group is the
algebra
$SDiff(T^2)$ of two-dimensional vector-fields preserving an area. The action
of this algebra on potential, $\tau$-function (and so on) in the
framework of the longwave limit of $2d$ Toda lattice was considered in
\cite{tak}.

The previous formulae can be used in the inverse direction. Let $E(p,x)$ and
$Q(p,x)$ be any solution of the equation (\ref{g8}). Using them
as Cauchy data for the equations (\ref{g2},\ref{g9} we define the functions
$E(p,T)$, $Q(p,T)$ that satisfy (\ref{g8}) for all $T$. After that
the potential $S(p,T)$ can be found with the help of formula
\beq
S(p,T)=\int^p Q(p,T)dE(p,T). \label{g10}
\eeq

Let us revise from this general point of view the definition of the
$\tau$-function corresponding to the solutions of the Whitham equations on
$\hat M_{0,N}$. As it was shown in the theorem 2.1 the local parameters
$k_{\alpha}$ by themselves are solutions of the equations (\ref{g2}).
Therefore,
they define a set of local potentials $S_{\alpha}(k_{\alpha})$ such that
the relation (\ref{33})
$$\Omega_B(k_{\alpha},T)=\partial_B S_{\alpha}(k_{\alpha},T),\ \ B=(\beta,i).$$
are fulfilled. On the other hand let us consider the solutions $E(p,T),Q(p,T)$
of the equations (\ref{g2}) with the initial data
\beq
E(p,x)=p,\ \ Q(p,x)=x. \label{g11}
\eeq
They are the holomorphic outside the punctures $p_{\alpha}(T)$. Hence,
there exists a ``global" differential $dS_0(p,T)$
that is holomorphic outside the punctures $p_{\alpha}(T)$, also. Let us define
a one-form on the space with the coordinates $t_A$
\beq
\delta \log \tau =\frac12 \sum_{\alpha=1}^N \bigl(\sum_{i=1}^{\infty}
{\rm res}_{\alpha}(k_{\alpha}^i dS_{\alpha}(p,T))dt_{\alpha,i}+
\frac{1}{2\pi i}\oint_{\sigma_{\alpha}} {\ln}(p-p_{\alpha})dS_0(p,T)
dt_{\alpha,0}\bigr),
\label{g12}
\eeq
where $\sigma_{\alpha}$ is a contour around the cut connecting $p_1=\infty$
and $p_{\alpha}$. It is easy to check with the help of the formulae
(\ref{40}-\ref{43}) that $\delta \log \tau $ is a closed form. Therefore,
locally there exist a $\tau$-function. What are advantages of the
algebraic solutions ?

As it was shown in section 3 for algebraic solutions there exist constants
$t_A^0$ such that a sum
\beq
S(p,T)=\sum_A (t_A-t_A^0)\Omega_A(p,T) \label{g13}
\eeq
is a ``global" potential coinciding in a neibourhoods of $p_{\alpha}$
with the local potentials $S_{\alpha}$. This provides the possibility
to define explicitle with the help of formula (\ref{35}) the $\tau$-function
but
not only its full external differential (\ref{g12}).

\section{Arbitrary genus case}

{\bf 1. Definition.}
The moduli space $\hat M_{g,N}$ is ``bigger" then $\hat M_{0,N}$. In the
 approach in which ``times" of the Whitham hierarchy are considered as a new
system of coordinates on the phase space it is natural to expect that there
should be more ``times" in the Whitham hierachy on $\hat M_{g,N}$. We shall
increase their number in a few steps. But at the begining let us consider
the {\it basic} Whitham hierachy in the form that  has arisen as a result
of the averaging procedure for the algebraic-geometrical solutions of
two-dimensional integrable equations. In this hierarchy there are the same
set of ``times" (\ref{(2.3)}) and this is the only part of universal Whitham
hierarchy on $\hat M_{g,N}$ that has a smooth degeneration to the zero-genus
hierarchy.

Let $\Gamma_g$ be a smooth algebraic curve of genus $g$ with local coordinates
$k_{\alpha}^{-1}(P)$ in neighbourhoods of $N$ punctures $P_{\alpha}$,
($k_{\alpha}^{-1}(P_{\alpha}) = 0$). Let us introduce meromorphic differentials

$1^0$. $d\Omega_{\alpha,i>0}$ is holomorphic outside $P_{\alpha}$ and has
the form
\beq
d\Omega_{\alpha,i}=d(k_{\alpha}^i+O(k_{\alpha}^{-1})) \label{701}
\eeq
in a neighbourhood of $P_{\alpha}$;

$2^0$. $\Omega_{\alpha,0},\ \alpha\neq1$ is a differential with simple poles at
$P_1$ and $P_{\alpha}$ with residues $1$ and $-1$, respectively
$$d\Omega_{\alpha,0}=dk_{\alpha}( k_{\alpha}^{-1}+O(k_{\alpha}^{-1}))$$
\beq
d\Omega_{\alpha,0}=-dk_1( k_1^{-1}+O(k_1^{-1}));\label{701'}
\eeq

$3^0$. The differentials $d\Omega_A$ are uniquelly normalized by the condition
that {\it all} their periods are real, i.e.
\beq
{\rm Im} \oint_c d\Omega_A=0,\ \  c\in H_1(\Gamma_g,Z). \label{702}
\eeq

The normalization (\ref{701}) does not depend on the
choice of basic cycles on $\Gamma_g$. Therefore, $d\Omega_A$ is indeed
defined by data $\hat M_{g,N}$.

Below, for the simplification of formulae we consider the complexcification of
the Whitham hierarchy on $\hat M_{g,N}$ that is a hierarchy on the moduli space
\beq
\hat M_{g,N}^{*}=\{\Gamma_g,P_{\alpha},k_{\alpha}^{-1}(P),\ \
a_i,b_i\in H_1(\Gamma_g,Z)\}, \label{703}
\eeq
where $a_i,\ b_i$ is a canonical basis of cycles on $\Gamma_g$, i.e. cycles
with the intersection matrix of the form $a_ia_j=b_ib_j=0,\ \
a_ib_j=\delta_{i,j}$. In this case the differentials $d\Omega_A$ should
be normalized by the usual conditions
\beq
\oint_{a_i}d\Omega_A=0,\ \ i=1,\ldots,g. \label{704}
\eeq
Both types of hierarchies  can be considered absolutely in parallel way.

Now we are going to show that generating equations (\ref{(1.6)}) in which
$\Omega_A$ are integrals of the above-defined differentials is equivalent to
a set of commuting evolution equations on $\hat M_{g,N}^{*}$ (or
$\hat M_{g,N}$, respectively). Let fix one point $P_1$ and choose as
a``marked" index $A_0=(1,1)$. The multi-valued function
\beq
p(P)=\Omega_{1,1} (P)=\int^P d\Omega_{1,1},\ \ P\in \Gamma_g \label{705}
\eeq
can be used as a coordinate on $\Gamma$ everywhere except for the points
$\Pi_s$, where $dp(\Pi_s)=0$. The parameters (\ref{(2.2)}), i.e.
$$\{p_{\alpha}=p(P_{\alpha}),\ v_{\alpha ,s},\ \alpha=1,\ldots,N,\
s=-1,0,1,2,\ldots\}$$
and additional parameters
\beq
\pi_s=p(\Pi_s), \ s=1,\ldots, 2g,\label{706}
\eeq
\beq
U_i^p= \oint_{b_i}dp,\  i=1,\ldots,g\label{707}
\eeq
are a full system of local coordinates on $\hat M_{g,N}^{*}$.
\begin{th}
The zero curvature-form (\ref{(1.16)}) of the Whitham hierarchy on
$\hat M_{g,N}^{*}$ is equivalent to the compatible system of evolution
equations
\beq
\partial_A k_{\alpha}(p,T)=\{k_{\alpha}(p,T),\Omega_A(p,T)\},\label{7017}
\eeq
\beq
\partial_A U_i^p=\partial_x U_i^A,\ \ {\rm where} \ \
U_i^A=\oint_{b_i}d\Omega_A,\label{7007}
\eeq
\beq
\partial_A \pi_s=\partial_A p(\Pi_s)=\partial_x\Omega_A(\Pi_s).\label{708}
\eeq
\end{th}
In \cite{D2} where the application of the Whitham equations for
generalized Landau-Ginsburg madels were considered for the first time it was
noticed that the construction of solutions of the Whitham equations
that were proposed by author in \cite{K4} can be reformulated
in the form that actually includes a new ``additional" flows commuting with
basic ones (\ref{708}). It have to be mentioned that only $g$ of them are
universal. Let us introduce a set of $g$ new times $t_{h,1},\ldots,t_{h,g}$
that
are coupled with normalized holomorphic dufferentials $d\Omega_{h,k}$
\beq
\oint_{a_i}d\Omega_{h,k}=\delta_{i,k},\ \ i,k=1,\ldots,g \ .\label{709}
\eeq
\begin{th} The basic Whitham hierarchy (\ref{708}) is compatible with
the system that is defined by the same equations but with new ``hamiltonians"
$d\Omega_{h,k}$.
\end{th}
The proof of the both theorems does not differ seriously from the usual
consideration in the Sato approach and we shall skip them.

\bigskip
\noindent{\bf 2. Algebraic orbits and exact solutions.}
Let us introduce a finite-dimensional subspaces of $\hat M_{g,N}^{*}$ that
are invariant with respect to the Whitham hierarchy. Consider a normalized
meromorphic differential $dE$ of the second kind (i.e. $dE$ has not residues at
any point of $\Gamma_g$) that has poles of orders $n_{\alpha}+1$ at
the points $P_{\alpha}$. ( Normalized means that
\beq
\oint_{a_i}dE=0.\label{710}
\eeq
for hierarchy on $\hat M_{g,N}^{*}$ and that $dE$ has real periods for
hierarchy
on $\hat M_{g,N}$.) The integral $E(p)$ of this differntial has the expansions
of the form
\beq
E(p)=p^n + u_{n-2} p^{n-2} + ... + u_0+O(p^{-1}),\label{711}
\eeq
\beq
E(p)=\sum_{s=1}^{n_{\alpha}} v_{\alpha,s}(p-p_{\alpha})^{-s}+O(1)\label{712}
\eeq
at the point $P_1$ and the points $P_{\alpha},\ \alpha\neq 1$, respectively.
The formula (\ref{d3}), i.e.
$$k_{\alpha}^{n_{\alpha}}(p)=E(p), \alpha=1,\ldots,N.$$
defines local coordinates $k_{\alpha}^{-1}$ in neighbourhoods of $P_{\alpha}$.
Therefore, we have defined the embedding of the moduli space
${\cal N}_g(n_{\alpha})$ of curves with fixed normalized  meromorphic
differential $dE$ into $\hat M_{g,N}^{*}$
\beq
{\cal N}_g(n_{\alpha})\subset  \hat M_{g,N}^{*}. \label{713}
\eeq
The dimension of this subspace equals
\beq
D=\dim {\cal N}_g(n_{\alpha})=3g-2+\sum_{\alpha=1}^N(n_{\alpha}+1).\label{714}
\eeq
There are two systems of local coordinates on ${\cal N}_g(n_{\alpha})$. The
first system is given by the coefficients of the expansions (\ref{711},
\ref{712})
\beq
\{u_i, i=0,\ldots,n-2;\ \ p_{\alpha},\ \
v_{\alpha,s}, s=1,\ldots, n_{\alpha})\},\label{715}
\eeq
and by the variables (\ref{706},\ref{707}), i.e.
$$\pi_s=p(\Pi_s), \ s=1,\ldots, 2g;\ \ U_i^p= \oint_{b_i}dp,\  i=1,\ldots,g.$$
The second system is given by the following parameters
\beq
U_i^E= \oint_{b_i}dE, \ \ i=1,\ldots,g, \label{716}
\eeq
\beq
E_s=E(q_s),\ \  {\rm where}\ \ dE(q_s)=0,\ \ s=1,\ldots,D-g.\label{717}
\eeq
Using the first system of coordinates it easy to show that
\begin{th}
The restriction of the Whitham hierarchy on ${\cal N}_g(n_{\alpha})$ is
given by the compatible system of equations (\ref{d5})
$$\partial_A E(p,T)=\{E(p,T),\Omega_A(p,T)\}.$$
\end{th}
(We whould like to remind that now besides
$t_{\alpha,i}$ the set of ``times" $t_A$ includes  the times $t_{h,k}$ that are

Let $dH_i$ be a  normalized differential that is defined on the cycle $a_i$,
i.e
\beq
\oint_{a_i} dH_i=0.\label{7171}
\eeq
For each set $H=\{dH_i\}$ of such differentials there exists a unique
differential $dS_H$ such that:

$dS_h$ is holomorphic on $\Gamma_g$ except for the cycles $a_i$ where
it has ``jumps"  that are equal to
\beq
dS_H^+(P)-dS_H^-(P)=dh_i(P),\ \ P\in a_i,\label{7172}
\eeq
\beq
\oint_{a_i}dS=0.\label{7173}
\eeq

\begin{th}
For any solution of the Whitham equations on ${\cal N}_g(n_{\alpha})$
there exist constants $t_A^0$ and constant differentials $dh_i$
(i.e. they do not depend on $T$) such that this solution
is given in an implicite form with the help of equations
\beq
{dS\over dp}(q_s,T)=0, \label{718}
\eeq
\beq
S(p,T)=\sum_A (t_A-t_A^0)\Omega_A(p,T)+dS_h.\label{719}
\eeq
\end{th}
The relations (\ref{718}) implies that
\beq
dS=QdE,\label{7191}
\eeq
where $Q(p)$ is holomorphic on $\Gamma_g$ outside the punctures $P_{\alpha}$
and
has ``jumps"
\beq
Q^+(E)-Q^-(E)=\frac{dH_i(E)}{dE},\ \ E\in a_i, \label{7192}
\eeq
on cycles $a_i$.

In this section we consider the solutions of the Whitham
hierarchy corresponding to the constant jumps, only, i.e.
\beq
dH_i(P)=t_{Q,k}dE(P). \label{7193}
\eeq
In that case $dQ$ is a single-valued differential on $\Gamma_g$. Let us present
an alternative formulation of the construction of such
solutions.

Consider the moduli space
\beq
\widetilde{\cal N}_g(n_{\alpha})=\{\Gamma_g,\ \ dQ,\ \ dE\}\label{724}
\eeq
of curves with fixed canonical basis of cycles, with fixed  normalized
meromorphic differential $dE$ havong poles of orders
$n_{\alpha}+1$ at points $P_{\alpha}$ and with fixed holomorphic outside the
punctures normalized differential $dQ(P)$.

The coordinates on this space are the variables (\ref{706},\ref{707},\ref{715})
$$\{\pi_s,\ \  U_i^p,\ \ u_i, \ \ p_{\alpha},\ \ v_{\alpha,s}, \}$$
and the coefficients of singular terms in the expansion
$$Q(p)=\sum_{j=1}^{\infty}b_{1,j}p^j+O(p^{-1}),$$
\beq
Q(p)=\sum_{j=1}^{\infty}b_{\alpha,j}(p-p_{\alpha})^{-j}+O(p-p_{\alpha}).
\label{720}
\eeq
The formulae (\ref{d9}), i.e.
\begin{eqnarray}
t_{\alpha,i}&=&\frac{1}{i}{\rm res}_{\alpha}(k_{\alpha}^{-i}(p)Q(p)dE(p))),
\ \ \  i>0; \nonumber\\
t_{\alpha,0}&=&{\rm res}_{\alpha}(Q(p)dE(p))\label{721}
\end{eqnarray}
and the formulae
\beq
t_{h,i}=\oint_{a_i}dS,\ \ i=1,\ldots,g,\ \ dS=QdE, \label{722}
\eeq
\beq
t_{Q,i}=-\oint_{b_i} dE, \ \ t_{E,i}=\oint_{b_i} dQ,\ \
i=1,\ldots,g.\label{725}
\eeq
defines times $t_A$ as functions on the space $\widetilde{\cal
N}_g(n_{\alpha})$

The differntials $d\Omega_{E,i},\ d\Omega_{Q,i}$ that are couple with times
$t_{E,i},\ t_{Q,i}$ are uniquelly defined with the help of following analytical

$1^0$. The differentials $d\Omega_{E,i},\ d\Omega_{Q,i}$ are holomorphic
on the curve $\Gamma_g$ everywhere except for the $a$-cycles, where they have
``jumps". Their boundary values on $a_j$ cycle satisfy the relations
$$ d\Omega_{E,i}^{+}-d\Omega_{E,i}^{-}=\delta_{i,j}dE,$$
\beq
d\Omega_{Q,i}^{+}-d\Omega_{Q,i}^{-}=\delta_{i,j}dQ;\label{725'}
\eeq
$2^0$.
\beq
\oint_{a_j}d\Omega_{E,i}=\oint_{a_j}d\Omega_{Q,i}=0,\ j=1,\ldots,g.\label{725"}
\eeq
In the same way as it was done in section 2 it can be shown that the number
of ``times" is equal to the dimension of $\widetilde{\cal N}_g(n_{\alpha})$.
Therefore, the ``times" $t_A$ can be considered as
new coordinates on $\widetilde{\cal N}_g(n_{\alpha})$, i.e.
\beq
\Gamma_g=\Gamma_g(T),\ dQ=dQ(T), \ dE=dE(T).\label{726}
\eeq
\begin{th}
For the differential
\beq
dS(E,T)=Q(E,T)dE\label{728}
\eeq
the following equalities
\beq
\partial_A S(E,T)=\Omega_A (E,T), \label{729}
\eeq
are fulfilled.
\end{th}
{\it Remark.} From the definition of times (\ref{721},\ref{722},\ref{725})
it follows
that
\beq
dS=\sum _{\alpha=1}^N \sum_{i=0}^{\infty} t_{\alpha,i}d\Omega_{\alpha,i}+
\sum_{k=1}^g t_{h,k}d\Omega_{h,k}+t_{E,k}\Omega_{E,k}.\label{729'}
\eeq
We shall give here a brief sketch of the proof (\ref{729}) for
$A= (Q,k)$, only, because for all other $A$ the proof is essentially the
same as for the proof of the theorem 2.2. Consider the derivative
$\partial_{Q,k} S(E,T)$. From the definition (\ref{728}) it follows that
$\partial_{Q,k} S(E,T)$ is holomorphic everywhere except for the
cycle $a_k$. On different sides of these cycle the coordinates are
$E^-$ and $E^+=E^--t_{Q,k}$. Hence, taking the derivative of the equality
\beq
Q(E^--t_{Q,k})-Q(E^-)=t_{E,k}, \ E^-\in a_k, \label{740}
\eeq
we obtain
\beq
\partial_{Q,k} S^+-\partial_{Q,k} S^-={dQ\over dE}. \label{741}
\eeq
Therefore, $\partial_{Q,k} S(E,T)=\Omega_{Q,k}$.

\begin{cor}
The integrals $E(p,T)$ and $Q(p,T)$ as functions of the variable
$p=\Omega_{1,1}$ satisfy the Whitham equations (\ref{d5}) and the classical
string equation (\ref{st}).
\end{cor}
(In both theorems a set of times $t_A$ includes all the times $t_{\alpha,i},
t_{h,i}, t_{E,i} , t_{Q,i}$.

\bigskip
\noindent{\bf 3. $\tau$-function}.
The $\tau$-function of the particular solution of the Whitham equation on
$\widetilde{\cal N}_g(n_{\alpha})$ that was constructed above are defined
by the formula
$$\ln \tau(T)=F(T),$$
\beq
F=F_0(T)+\frac{1}{4\pi i}\sum_{k=1}^g\oint_{a_k^-} t_{E,k} E dS
-\oint_{b_k} t_{h,k}dS+t_{h,k}t_{E,k}E_k,\label{742}
\eeq
where $F_0(T)$ is given by (\ref{35}), i.e.
$$F_0={1\over 2}\sum_{\alpha=1}^N {\rm res}_{\alpha}
(\sum_{i=1}^{\infty}  t_{\alpha,i} k_{\alpha}^idS(p,T))+
t_{\alpha,0}s_{\alpha}(T),$$
(the first integral in (\ref{742}) is taken over the left side of the $a_k$
cycle and $E_k=E(P_k)$ where $P_k$ is the intersection point of $a_k$ and $b_k$
cycles).

{\it Remark.} The differential $dS$ is discontinuous. Therefore, its
integral over $b_k$-cycle depends on the choice of the cycle. The last
term in (\ref{742}) restors an invariance (i.e. $F$ depends on the homology
class of cycles, only).

\begin{th}
For the above-defined $\tau$-function the equalities (\ref{37},
\ref{38})
are fulfilled. Besides this,
\beq
\partial_{h,k} F=\frac1{2\pi i}(t_{E,k}E_k-\oint_{b_k}dS),\label{743}
\eeq
\beq
\partial_{E,k} F=\frac1{2\pi i}(\oint_{a_k}EdS),\label{744}
\eeq
\beq
\partial_{Q,k} F=\frac1{4\pi i}(\oint_{a_k}QdS-2t_{E,k}t_{h,k}).\label{745}
\eeq
\end{th}
The proof of all these equalities is analogous to the proof of (\ref{37},
\ref{38}) and use different types of identities that can be proved with the
help of usual considerations of contour integrals.

\begin{cor}
For $A=(\alpha,i)$ the second derivatives $\partial_{A,B}^2 F$ are given by
the formulae (\ref{45},\ref{46}). Besides this,
\beq
\partial_{(h,k);A}^2F=\frac1{2\pi i}(E_k\delta_{(E,k);A}+Q_k\delta_{(Q,k);A}-
\oint_{b_k}d\Omega_A),\label{7451}
\eeq
\beq
\partial_{(E,k);A}^2=\frac1{2\pi i}(\oint_{a_k}Ed\Omega_A),\label{7441}
\eeq
\beq
\partial_{(Q,k);A}^2 F=\frac1{2\pi i}(\oint_{a_k}Qd\Omega_A-\partial_A
(t_{E,k}t_{h,k})).\label{7442}
\eeq
\end{cor}
We would like to mention that in particular the formula (\ref{7451}) gives
a matrix of $b$-periods of normalized holomorphic differentials on $\Gamma_g$
\beq
\partial_{(h,i);(h,j)}^2 F=-\oint_{b_i}d\Omega_{h,j}.\label{746}
\eeq
( for the particular case this relation for the first time was obtained
in \cite{D2}).

\begin{th}
The third derivatives of F(T) are equal to
\beq
\partial_{ABC}^3 F=\sum_{q_s} {\rm res}_{q_s}
\bigl({d\Omega_A d\Omega_B d\Omega_C \over dQ\  dE} \bigr)+\eta_{ABC},
\label{7461}
\eeq
where
$$\eta_{ABC}=0 \ \ {\rm if}\ \  A,B,C\neq (Q,k),$$
$$\eta_{AB(Q,k)}=\frac{1}{2\pi i}
\oint_{a_k}{d\Omega_A d\Omega_B\over dE}\ \ {\rm if}\ \ A,B\neq Q,k).$$
\end{th}

\bigskip
\noindent{\bf 4. Virasoro constraints.}
In this subsection we present ``$L_0,\ L_{-1}$" constraints for the
$\tau$-function of the homogenious solution of the Whitham hierarchy on
$\widetilde{\cal N}_g(n_{\alpha})$.

Consider the differential $Q^2dE$. It is holomorphic on $\Gamma_g$ outside
the punctures and cycles $a_k$ where it has jumps
\beq
(Q^2dE)^+-(Q^2dE)^-=2t_{E,k}QdE=2t_{E,k}dS.\label{747}
\eeq
Therefore,
\beq
\sum_{\alpha=1}^N {\rm res}_{\alpha}(Q^2dE)+
\frac 1{\pi i} \sum_{k=1}^g t_{E,k}t_{h,k}=0.\label{748}
\eeq
The expansion of $Q$ near the puncture $P_{\alpha}$ has a form (\ref{v6}).
Its substitution into (\ref{748}) gives
$$
\sum_{\alpha=1}^N \frac{1}{n_{\alpha}} \bigl(
\sum_{i=n_{\alpha}+1}^{\infty} it_{\alpha,i}
\partial_{\alpha,i-n_{\alpha}} F +
n_{\alpha}t_{\alpha,0}t_{\alpha,n_{\alpha}}+
{1\over 2}
\sum_{j=1}^{n_{\alpha}-1} j(n_{\alpha}-j)t_{\alpha,j}t_{\alpha,n_{\alpha}-j}
\bigr)
$$
\beq
+\frac 1{2\pi i} \sum_{k=1}^g t_{E,k}t_{h,k}
=0. \label{749}
\eeq
In the same way the consideration of the differential $Q^2EdE$ proves an
analogue of $L_{-1}$ constraint:
\beq
\sum_{\alpha=1}^N \frac{1}{n_{\alpha}} \sum_{i=1}^{\infty}it_{\alpha,i}
\partial_{\alpha,i} F+\frac 1{2\pi i} \sum_{k=1}^g t_{E,k}\partial_{E,k} F
+\frac12 t_{\alpha,0}^2=0;
\label{750}
\eeq

{\it Remark.} In order to obtain higher ``$L_{n>0}$" Virasoro constraints
one has introduce $p$-gravitational decendants of the
``fields" $d\Omega_{E,k}$ that are holomorphic differentials on $\Gamma$ except

\bigskip
\noindent{\bf 5. Landau-Ginzburg type models on Riemann surfaces}
In this subsection we present a generalisation of the results of section 5
for the case of Riemann surfaces of an arbitrary genus.
Let us consider a genus $g$ Riemann surface $\Gamma_g$ with fixed
canonical basis of cycles and with fixed meromorphic normalized
differential $dE$, i.e. a point of the moduli space ${\cal N}_g(n_{\alpha})$.
The same formulae (\ref{T8},\ref{T9}), as in genus zero case, define
a Fr\"obenius algebra ${\cal R}_{\Gamma_g,dE}$
\beq
{\cal R}_{\Gamma_g,dE}=\hat{\cal R}/(dE=0),\label{l1}
\eeq
where  $\hat{\cal R}$ is a ring of all meromorphic functions that are regular
at the zeros $q_s$ of the differential $dE$. The formula
\beq
\langle f,g\rangle=\sum_{q_s} {\rm res}_{q_s} \bigl( {f(p)g(p)\over E_p}dp
\bigr),\ f(p),\ g(p) \subset \hat{\cal R},\label{l2}
\eeq
defines a non-degenerate scalar product on ${\cal R}_{\Gamma_g,dE}$.

For any g a ``small phase" space is a space of times $t_a$ with
indecies $a\in{\cal A}_{sm}^g$, where ${\cal A}_{sm}^g$ is a union of
${\cal A}_{sm}$ (that was defined in section 5)
$${\cal A}_{sm}=\{ a=(\alpha,i)| \alpha=1,\ i=1,\ldots, n-1;\
\alpha=2,\ldots,N,\ i=0,\ldots,n_{\alpha} \}$$
and indecies $(h,k),(E,k)$. In the basis
\beq
\phi_a={d\Omega_a\over dp}\label{l3}
\eeq
the scalar product has the form:
\beq
\langle \phi_a \phi_b \rangle=\eta_{ab}={[i][j]\over n_{\alpha}}
\delta_{\alpha,\beta} \delta_{i+j,n_{\alpha}}, \ \  a,b \in{\cal A}_{sm}
\label{l4}
\eeq
\beq
\langle \phi_{E,k} \phi_{h,s} \rangle=\delta_{k,s} \label{l5}
\eeq
otherwise zero (here $[i]$ is the same as in (\ref{v13}).

Let us consider the Whitham ``times" $t_a$ that were defined in
(\ref{721},\ref{722},\ref{725}) for the choice $dQ=dp$, i.e.
\begin{eqnarray}
t_{\alpha,i}&=&\frac{1}{i}{\rm res}_{\alpha}(k_{\alpha}^{-i}(p)pdE(p))),
\ \ \  \alpha,i>0\in {\cal A}_{sm}; \nonumber\\
t_{\alpha,0}&=&{\rm res}_{\alpha}(pdE(p));\label{l6}\\
t_{h,k}&=&\oint_{a_k}pdE,\ \ ,k=1,\ldots,g;\label{l7}\\
t_{p,i}&=&-\oint_{b_k} dE, \ \ t_{E,k}=\oint_{b_k} dp,\ \ k=1,\ldots,g.
\label{l8}
\end{eqnarray}
\begin{th}
A Jacobian of the map
\beq
{\cal N}_g(n_{\alpha})\ \longmapsto \ \{t_a,\ a\in{\cal A}_{sm}^g \}
\label{l8a}
\eeq
is nonvanishing anywhere (i.e. the map (\ref{l8a}) is a non-remified covering
of some domain of the complex space with coordinates $t_a$).
\end{th}.

Let us fix the values $t_{p,k}=t_{p,k}^0$ and consider the restriction of
$F=\ln \tau$ on the affine space that is ${\cal A}_{sm}^g$ shifting
by
$$t_{1,n+1}={n\over n+1}, \ \ t_{p,k}=t_{p,k}^0.$$
Then from the statement of the theorem 7.7 it follows that
\beq
\partial_{abc}^3 F=\sum_{q_s} {\rm res}_{q_s} \bigl(
{d\Omega_a d\Omega_b d\Omega_c \over dp\  dE} \bigr), \
a,b,c\in {\cal A}_{sm}^g.\label{l9}
\eeq
\begin{cor}
The dependence of Fr\"obenius algebrae corresponding to ${\cal
N}_g(n_{\alpha})$
with respect to the coordinates $t_a,\ \ a\in {\cal A}_{sm}^g$ is
a potential deformation.
\end{cor}
In \cite{D2} the particular case of this statement was proved. It corresponds
to the Whitham hierarchy on moduli space of genus $g$ curves with
fixed function $E(P)$ having a pole of order $n$ at a point $P_1$.
(This moduli space is a subspace of ${\cal N}_g(n)$ that is specified
by the conditions $t_{p,k}=0$.) The differential-geometrical interpretation
of Whitham coordinates that was proposed in \cite{D2} is valid in general case
as well.

Let us denote a subspace of ${\cal N}_g(n_{\alpha})$ corresponding to fixed
values of $t_{p,k}=t_{p,k}^0$ by ${\cal N}_g(n_{\alpha}|t_{p,k}^0)$. A
system of local coordinates on its open submanifold ${\cal D}$ is given by
(\ref{717}), i.e.
$$E_s=E(q_s),\ \  {\rm where}\ \ dE(q_s)=0,\ \
s=1,\ldots,D-g=2g-2+\sum_{\alpha=1}^N(n_{\alpha}+1).$$
Submanifold ${\cal D}$ can be defined as a submanifold on which the values
$E_s$ are distinct. The formula
\beq
ds^2=\sum_{s=1}^{D-g} {\rm res}_{q_s}({dpdp\over dE})(dE_s)^2\label{l10}
\eeq
defines a metric on ${\cal D}\subset{\cal N}_g(n_{\alpha}|t_{p,k}^0)$.
The scalar products of the vector-fields $\partial_a={\partial \over
\partial t_a}$ with respect to the metric (\ref{l10}) have the form:
\beq
\langle \partial_a \partial_b \rangle=\eta_{ab}={[i][j]\over n_{\alpha}}
\delta_{\alpha,\beta} \delta_{i+j,n_{\alpha}}, \ \  a,b \in{\cal A}_{sm}
\label{l11}
\eeq
\beq
\langle \partial_{E,k} \partial_{h,s} \rangle=\delta_{k,s}, \label{l12}
\eeq
othewise zero. The proof of (\ref{l11},\ref{l12}) is based on the
formula (\ref{r9})
\beq
\partial_A E_s={d\Omega_A \over dp}. \label{l13}
\eeq
and formulae (\ref{l4},\ref{l5}). Consequently, in the Whitham coordinates
$t_a, \ a\in {\cal A}_{sm}^g$ the metric (\ref{l10}) has
a constant coefficients (i.e. $ds^2$ is a flat metric and $t_a$ are
flat coordinates). In \cite{D2} the formulae (\ref{l11},\ref{l12}) and
the following two main
arguments were used for the proof that the functions $\{t_a, a\in{\cal A}_{sm},
t_{E,k},t_{h,k}\}$ define a system of coordinates everywhere on the
moduli space of genus $g$ curves with fixed function $E(P)$ having
a pole of order $n$ at a point $P_1$. First of all, the functions
$t_a, a\in{\cal A}_{sm}, t_{E,k},t_{h,k}$ are holomorphic functions on
${\cal N}_g(n_{\alpha}|t_{p,k}^0)$. The second argument that had been used is:
the ``dual" metric
\beq
d\hat s^2=\sum_{s=1}^{D-g} {\rm res}_{q_s}
({dpdp\over dE})^{-1}({\partial \over \partial E_s})^2\label{l14}
\eeq
on the cotangent bundle can be extended as a {\it smooth} on the whole
moduli space. This very powerfull statement is a particular case of general
results that were obtained by Novikov and Dubrovin in the framework of their
Hamiltonian approach for the Whitham theory \cite{ND}. We would like to
mention that the last argument can be replaced by the corollary of the
theorem 7.1, because the Sato-form (\ref{7017},\ref{7007},\ref{708}) shows
that {\it vector-fields} $\partial_a$ are {\it smooth} on the whole
moduli space.

\bigskip
{\bf Acknowledgments.} The author would like to thank E.Bresan, J.-L. Gervais,
V.Kazakov, B.Dubrovin for many useful discussions. He also wishes to thank
Laboratoire de Physique Th\'eorique de l'\'Ecole Normale Sup\'erieure for
kind hospitality during period when this work was done.


\begin{thebibliography}{**}

\bibitem {BK..}
F.Bresin, V.Kazakov, {\it Phys Lett.} {\bf B 236} (1990) 144.

M.Douglas, S. Shenker, {\it Nucl.Phys.} {\bf B 335} (1990) 635.

D.J.Gross, A. Migdal, {\it Phys.Rev.Lett.} {\bf 64} (1990) 127.

D.J.Gross, A. Migdal, {\it Nucl. Phys.} {\bf B 340} (1990) 333.

T.Banks, M.Douglas, N.Seiberg, S.Shenker, {\it Phys. Lett.} {\bf B 238}

(1990) 279.

\bibitem{npergr}
V.Kazakov, {\it Phys.Lett.} {\bf 159 B} (1985) 303.

F.David, {\it Nucl. Phys.} {\bf B 257} (1985) 45.

V.Kazakov, I.Kostov, A.Migdal {\it Phys. Lett.} {\bf 157 B} (1985) 295.

J.Fr\"olich, {\it The statistical mechanics of surfaces} in
{\it Applications of Field Theory to Statistical Mechanics}, L.Garrido ed.
(Springer,1985).

\bibitem{W1}
E.Witten, {\it Nucl.Phys} {\bf B 340} (1990) 281.

\bibitem{W2}
E.Witten, {\it Two-dimensional gravity and intersection theory on
moduli space} , Surveys In Diff. Geom. {\bf 1} (1991) 243.

\bibitem{TopG}
J.Labastida, M.Pernici, E.Witten, {\it Nucl.Phys} {\bf B310} (1988) 611.

D.Montano, J.Sonnenschein, {\it Nucl.Phys} {\bf B 313} (1989) 258;
{\it Nucl.Phys} {\bf 324} (1990) 348.

R. Myers, V. Periwal, {\it Nucl.Phys} {\bf 333} (1990) 536.

\bibitem{Kon}
M.Kontsevich, {\it Funk.Anal. i Pril} {bf 25} (1991) 50.

M.Kontsevich {\it Intersectuion theory on the moduli space of curves
and matrix Airy function} Max-Planck-Institute preprint MPI/91-47.

\bibitem{K1}
I.Krichever, {\it Comm.Math.Phys.} (1991) 1.
\bibitem{K2}
I.Krichever, {\it Whitham theory for integrable systems and topological
field theories} (to appear in procedings of {\it Summer Cargese School,
July,1991}

\bibitem{VVD}
 E. Verlinder, H. Verlinder, {\it A solution of two-dimensional topological
quantum gravity, preprint IASSNS-HEP-90/40, PUPT-1176 (1990)}.

\bibitem{FK}
M. Fukuma, H. Kawai, Continuum Schwinger-Dyson equations
and universal structures in two-dimensional quantum gravity ,
preprint Tokyo University UT-562 , May 1990 .

M. Fukuma, H. Kawai, Infinite dimansional Grassmanian structure
of two-dimensional quantum gravity , preprint Tokyo University UT-572 ,
November 1990.


\bibitem{D2}
B.Dubrovin, {\it Hamiltonian formalism of Whitham-type hierarchies and
topological Landau-Ginsburg models}, Preprint 1991,
(Submitted to {\it Comm.Math.Phys.}).

\bibitem{ND}
B.Dubrovin, S.Novikov, {\it Sov.Math.Doklady} {\bf 27} (1983) 665.

S.Novikov, {\it Russ.Math Surveys} {\bf 40}:4 (1985) 85.

B.Dubrovin, S.Novikov, {\it Russ.Math.Surveys} {\bf 44}:6 (1989) 35.

B.Dubrovin, {\it Geometry of Hamiltonian Evolutionary Systems}, Bibliopolis,
Naples 1991.

\bibitem{D3}
B.Dubrovin, {\it Integrable systems in topological field theory},
prepint INFN-NA-A-IV-91/26, Napoly, 1991.

\bibitem{K3}
I.M.Krichever, {\it Doklady Acad. Nauk USSR} {\bf 227} (1976) 291.

I. Krichever, {\it Funk. Anal. i Pril.} {\bf 11} (1977)  15.

\bibitem{NDMI}
B. Dubrovin, V. Matveev, S. Novikov, {\it Uspekhi Mat. Nauk} {\bf 31}:1
(1976) 55-136.

V. Zakharov, S. Manakov, S. Novikov, L. Pitaevski,{\it Soliton theory}
Moscow, Nauka, 1980.

\bibitem{K4}
I. Krichever, {\it Funk. Anal. i Pril.} {\bf 22}(3) (1988)  37-52.

\bibitem{K5}
I. Krichever, {\it Uspekhi Mat. Nauk} {\bf 44}:2 (1989) 121.

\bibitem{WFMD}
A. Gurevich, L.Pitaevskii, {\it JETP} {\bf 65}:3 (1973), 590.

H. Flashka, M. Forest, L.McLaughlin, {\it Comm. Pure and Appl. Math.} \ \ \
{\bf 33}:6.

S. Yu. Dobrokhotov, V. P. Maslov , {\it Soviet Scientific Reviews, Math. Phys.
Rev. OPA Amsterdam} {\bf 3} (1982)  221-280.

\bibitem{tak}
K.Takasaki, K.Takebe, {\it SDiff(2) Toda equation-hierarchy, tau-function
and symmetries}, preprint RIMS-790, Kyoto.

K.Takebe, {\it Area-Preserving Diffeomorphisms and Nonlinear Integrable
Systems}, in proceedings of {\it Topological and geometrical methods in
field theory}, May 1991, Turku, Finland.

\bibitem{Z}
V.Zakharov , {\it Funk. Anal. i Pril.} {\bf 14} (1980) 89.

\bibitem{Tzar}
S.Tsarev, {\it Izvestiya USSR, ser. matem} (1990).

\bibitem{sav}
M.Saveliev, {\it On the integrability problem of the continuous
long wave approximation of the Toda lattice}, preprint ENSL, Lyon, 1992.

\bibitem{mak}
Yu. Makeenko, {\it Loop equations in matrix models and in 2D quantum gravity}
(Submitted in {\it Mod. Phys. Lett.A}).

\bibitem{dug}
M.Douglas, {\it Phys.Lett.} {\bf B238} (1990) 176.

\bibitem{Eg}
 T. Eguchi, S.-K. Yang, {\it N=2 superconformal models as topological
field theories }, preprint of Tokyo University  UT-564  (1990).

\bibitem{Li}
K.Li, {\it Topological gravity with minimal matter} , Caltech-preprint
CALT-68-1662 .

\bibitem{vafa}
E. Martinec, {\it Phys. Lett.}  {\bf 217B} (1989),  431.

C. Vafa, N. Warner, {\it Phys. Lett.} {\bf 218B} (1989),  51.

W. Lerche, C. Vafa, N.P. Warner, {\it Nucl. Phys.} {\bf  B324} (1989),  427.

\bibitem{D1}
B.Dubrovin, {\it Differential geometry of moduli spaces and its application to
soliton equations and to topological conformal field theory}, preprint
No 117 of Scuola Normale Superiore, Pisa, November 1991.

\end{thebibliography}
\end{document}